\documentclass[10pt,review]{elsarticle}

\usepackage{lineno, hyperref}
\modulolinenumbers[5]

\usepackage[a4paper]{geometry}
\journal{Sustainable Cities and Society}
\usepackage{bigstrut}
\usepackage{graphicx}
   \graphicspath{{figure_eps/}}              
\usepackage{amsmath}               
\usepackage{siunitx}
\usepackage{caption}
\usepackage{subcaption}
\usepackage{float} 
\usepackage{tabularx}
\usepackage{tabu}
\usepackage{csquotes}

\usepackage{colortbl}
 \usepackage{blindtext}
    \usepackage{multicol}
    \usepackage[british,UKenglish,english]{babel}
\usepackage{titlesec} 
\usepackage{color,soul}
\usepackage{booktabs}
\newcommand{\tabitem}{~~\llap{\textbullet}~~}

\usepackage{siunitx}
\usepackage{enumerate}
\usepackage{siunitx}
\usepackage{breakurl}
\usepackage{epstopdf}
\usepackage{pbox}
\usepackage{multirow}
 \usepackage{blindtext} 
 \usepackage{enumitem}
\usepackage{epstopdf}

\begin{document}

\begin{frontmatter}
\title{High-Level Penetration of Renewable Energy with Grid: Challenges and Opportunities} 
\author{Md Shafiul Alam$^{{1},\ast}$, Fahad Saleh Al-Ismail$^{1,2}$, M. A. Abido$^{1,2}$,  and Aboubakr Salem$^2$}
\address{$^1$K.A.CARE Energy Research \& Innovation Center, King Fahd University of Petroleum \& Minerals, 
Dhahran, Saudi Arabia\\
$^2$Department of Electrical  Engineering, King Fahd University of Petroleum \& Minerals, 
Dhahran, Saudi Arabia\\
}
\cortext[mycorrespondingauthor]{Corresponding author}
\ead{mdshafiul.alam@kfupm.edu.sa or shafiulmsc\_ buet@yahoo.com}

\begin{abstract}
 The utilization of renewable energy sources (RESs) has become significant throughout the world especially over the last two decades.
 Although high-level RESs penetration  reduces negative environmental impact compared to conventional fossil fuel based energy generation, control issues become more complex as well as total inertia to the system is significantly decreased due to removal of conventional synchronous generators. Some other technical issues, high uncertainties, low fault ride through capability, high fault current, low generation reserve, and low power quality, arise due to RESs integration. Renewable energy like solar and wind are highly uncertain due to intermittent nature of wind and sunlight. Cutting edge technologies including different control strategies, optimization techniques, energy storage devices, and fault current limiters are employed to handle those issues. This paper summarizes several challenges in the integration process of high-level RESs to the existing grid. The respective solutions to each challenge are also discussed. A comprehensive list of challenges and opportunities, for both wind and solar energy integration cases, are well documented. Also, the future recommendations are provided to solve the several problems of renewable integration which could be key research areas for the industry personnel and researchers. 

\end{abstract}


\end{frontmatter}

$ ~$\\
\textbf{Abbreviations}\\
The following abbreviations are used in this manuscript: \\

\noindent 
\begin{tabular}{@{}ll}
RESs  & Renewable Energy sources~~~~~~~~~~~~~~~~~~~~~~~~\\
FRT & Fault ride through\\
PFC & Primary frequency control \\
DGs & Distributed generations \\
PCC  & Point of common coupling\\
PV   & Photovoltaic \\
ROCOF  & Rate of change of frequency\\
ESSs & Energy storage systems\\

\end{tabular}
\begin{tabular}{@{}ll}
VSC & Voltage source converter\\
PLL & Phase locked loop\\
AGC & Automatic generation control\\
PMSG & Permanent magnet synchronous generator\\
BESS & Battery energy storage system\\
SMES & Superconducting magnetic energy storage\\
LVRT & Low voltage ride through\\
SCIG  & Squirrel case induction generator\\
\end{tabular}

\begin{tabular}{@{}ll}
FACTS  & Flexible alternating current transmission system\\
FLC  & Fault current limiter\\
HVDC  & High voltage direct current\\
SDBR  &  Series dynamic braking resistor~~~~~~~~~~~~~~~~~~~\\
BFCL  & Bridge fault current limiter \\
STATCOM  & Static synchronous compensator\\
SVC & Static var compensator\\
TCSC & Thyristor controller series capacitor~~~~~~~~~\\
UPFC & Unified power flow controller\\
UPQC & Unified power quality conditioner\\
PFs & Passive filters\\
SAPF & Shunt active power filter\\

\end{tabular}
\begin{tabular}{@{}ll}
HTS & High temperature superconducting\\
LMMN & Least mean mixed norm\\
DVR & Dynamic voltage restorer\\
WTPC & Wind turbine power curve\\
UC & Unit commitment\\
ED & Economic dispach \\
IGDT & Information-gap decision theory\\
GA & Genetic Algorithm\\
PSO & Particle swarm optimization\\
DE & Differential evolution\\
ANN & Artificial neural network\\
MAE & Mean absolute error\\

\end{tabular}

\section{Introduction}
\begin{figure}[H]
\begin{center}

 \includegraphics[scale=1.2]{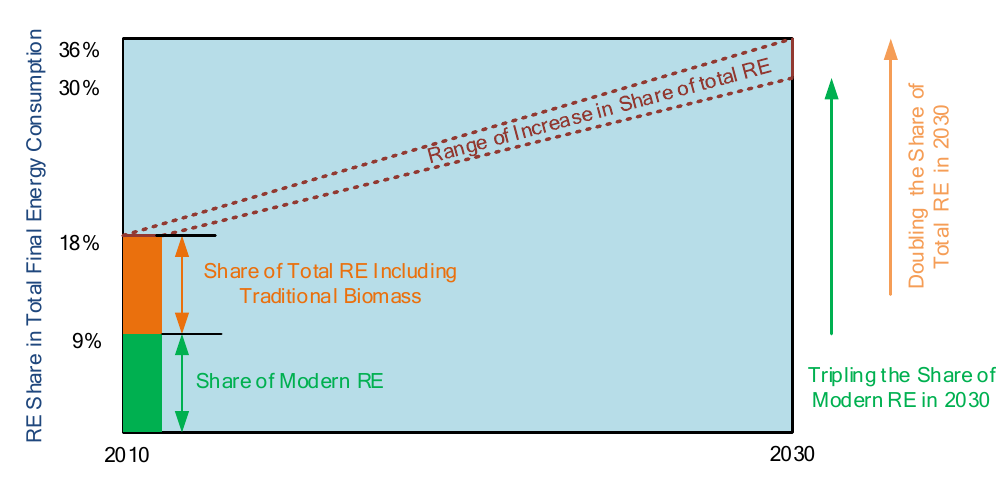}
\caption{A roadmap for renewable power generation by 2030}
\label{fig:sper}
\end{center}
\end{figure}

Nowadays, several environmental concerns arise due to emission of carbon dioxide, Sulphur dioxide, and nitrogen oxide by the fossil fuel based generating stations. This environmental pollution causes global warming and acid rain\cite{warner201921st}. On the other hand, renewable energy (RE) generation systems are clean and cheaper as compared to traditional synchronous generator based power generations. Thus, the governments and several agencies are forced to increase the renewable energy generation to replace fossil fuel based power generation. As per international renewable energy agency \cite{irena2014remap}, a roadmap for RE integration in the world upto 2030  is shown in Figure \ref{fig:sper}. It is expected that the world will meet  total 36\% of its energy demand from renewable energy sources (RESs) by 2030.    
Solar, wind, tide, wave, and geothermal heat are the main sources of renewable energy  generation. Among this sources solar and wind systems are most promising due lower cost of generation and maximum power point tracking capability over a wide range of wind and sunlight variation \cite{hossain2016transient,hossain2015transient}.

\begin{figure}
        \centering
        \begin{subfigure}[h]{0.48\textwidth}
                \includegraphics[width=\textwidth]{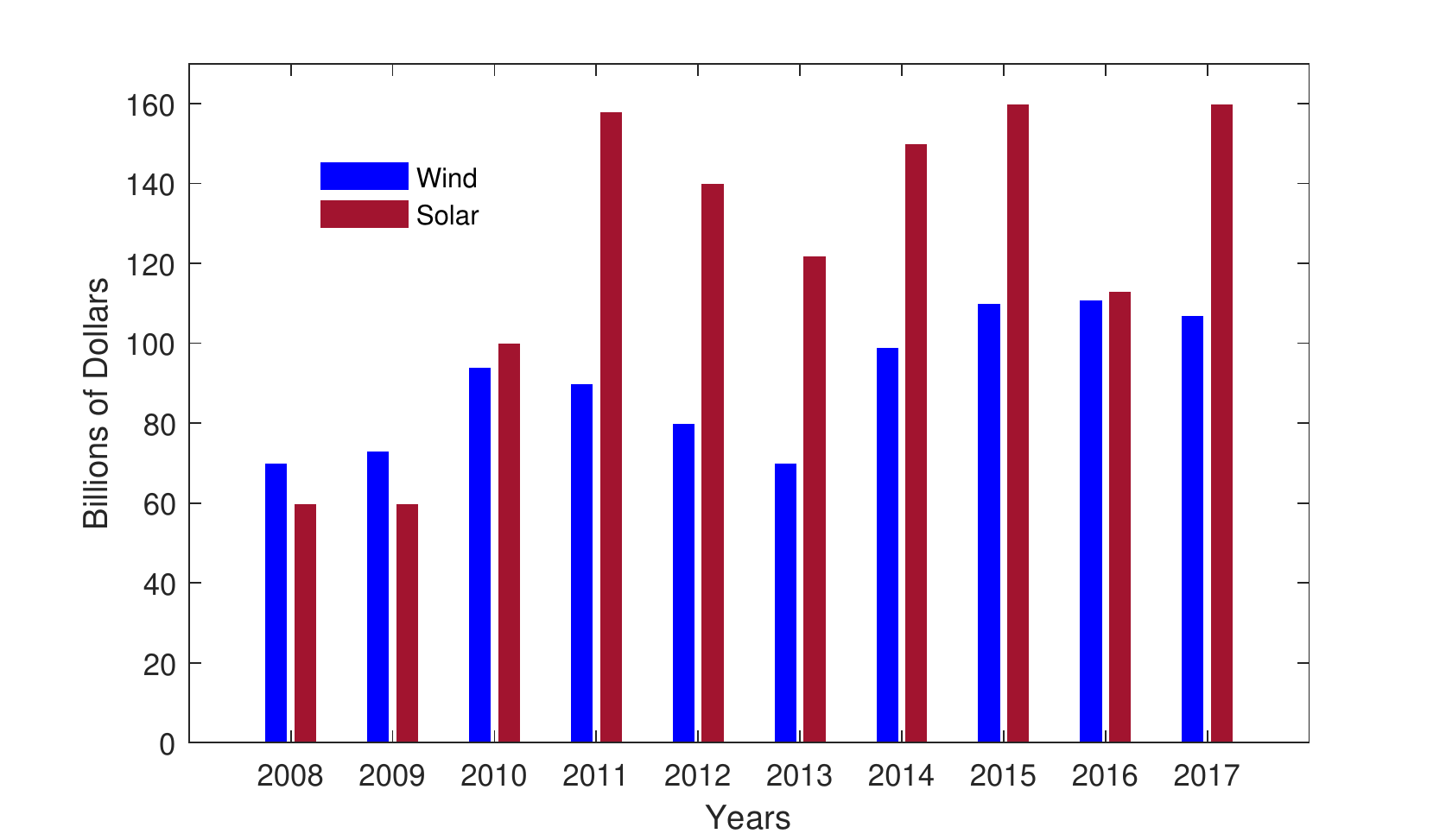}
                \caption{ }
                \label{fig:f2_a}
        \end{subfigure}
        \quad%
         \begin{subfigure}[h]{0.480\textwidth}
                \includegraphics[width=\textwidth]{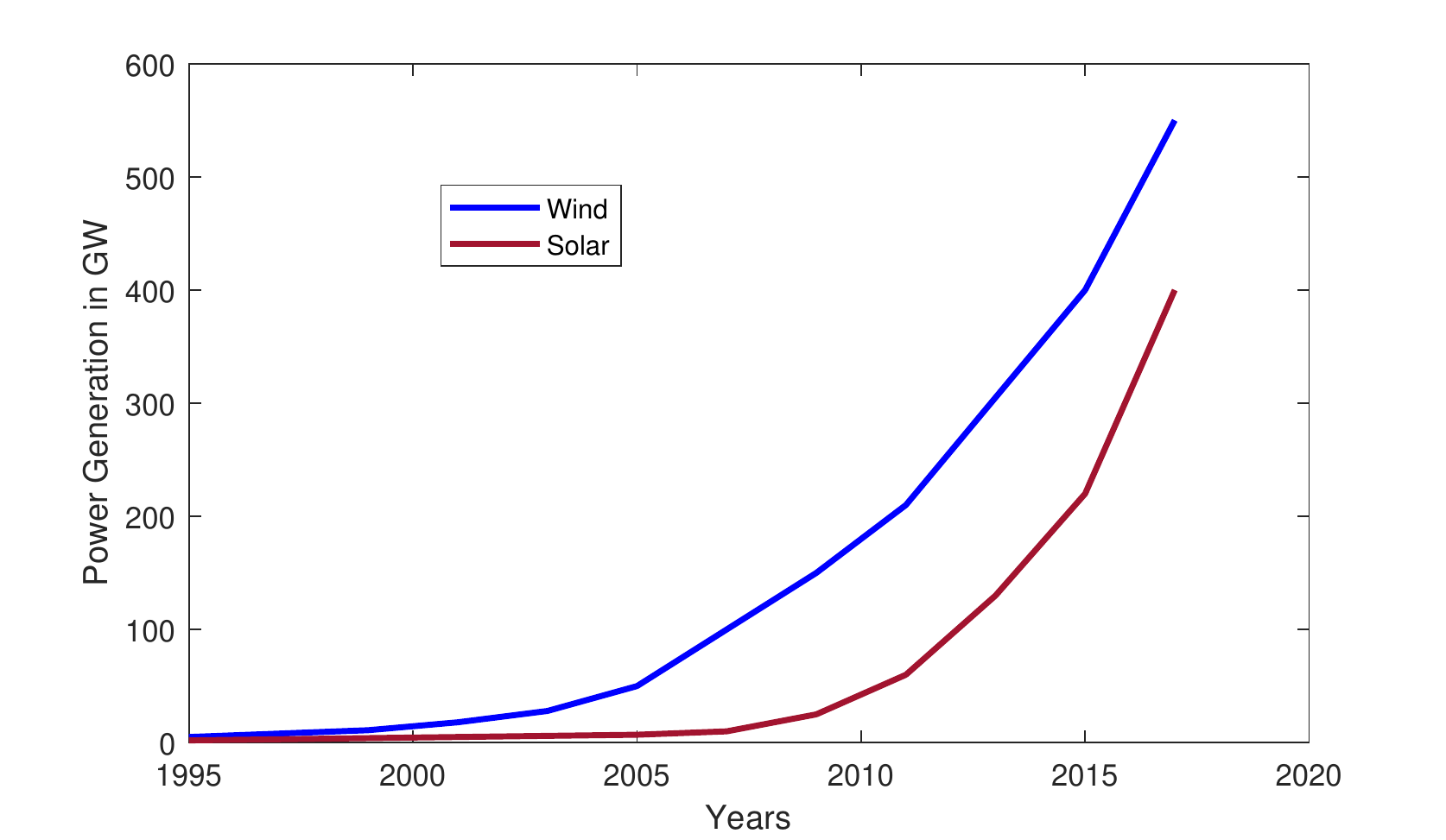}
                \caption{ }
                \label{fig:f2_b}
        \end{subfigure}
       \vspace{-10pt}
                  \caption{Wind and solar power (a) world wide investment (b) power generation}
        \label{fig: solarwind}
\end{figure}

Figure \ref{fig: solarwind} shows the global investment and generation of power from wind and solar energy resources. As shown in Figure \ref{fig:f2_a}, more money was invested for power generation from wind until 2009. However, this scenario was reversed, since then\cite{world2018,hossain2019evolution}. The global power generation in giga watt (GW) from wind and solar is depicted in Figure \ref{fig:f2_b}.
 
The high-level integration of RE to the utility grid may lead to concerns regarding stable and reliable operation of the system due to stochastic nature of power generation \cite{hossain2017overview}. This is because of continuous wind speed and sunlight irradiation variations with time. The intermittent and unpredictable nature of renewable energy sources could be modeled properly to reduce the negative impact on stable operation of the system. Several methodologies\cite{wang2017uncertain,kumar2017recent,athari2016modeling} are presented in the literature to model uncertainties in RE in order to have minimal impact on reliable and stable operation of the system.    
The proper control of the power electronic (PE) converters connected to the renewable energy sources is important to ensure the stable operation during the transients and AC system parameter variations. As per grid code requirement, renewable energy sources should stay connected during system faults. Thus, improvement of fault ride through (FRT) capability of renewable energy conversion system becomes vital. Several methods have been proposed in the literature to improve the FRT capability of solar and wind energy systems connected to the grids\cite{rashid2014bridge,naderi2017low}.    
Renewable energy conversion systems employ costly PE converters for power conditioning. Protection of such converters is important from both economic and stability point of view. However, short-circuit power level increases with the increased level of renewable energy sources integration. Thus, in order to restrict the fault current within acceptable limits, several control strategies without and  with auxiliary devices, such as fault current limiter, energy storage device, dynamic voltage restorer, are presented in the literature\cite{naderi2017low,xiao2017enhancing,daoud2015flywheel,fereidouni2012impact}.  

Uncertainty in renewable energy generation creates several problems like supply-demand mismatch and reserve generation reduction, posing frequency instability problem in the system. Also, large-scale RESs integrated system faces extremely low inertia which further degrades system frequency stability. The concept of virtual synchronous generator has evolved which imitates the behavior of prime mover to enhance inertia in the control loop virtually and accordingly stabilizing the system frequency \cite{dreidy2017inertia,magdy2019renewable}.  A central control scheme is presented in \cite{pourmousavi2012real}  to incorporate loads in primary frequency control (PFC). However, the control scheme fully depends on fast communication link which may pose threat due to cyber-attack. In order to avoid the need for fast communication channel, different schemes are proposed in the literature for implementing local controllers for loads. However, for improvement of the local load controller, some parameters need to be determined in the main control center and transferred to the local controller through communication links. In \cite{molina2010decentralized}, loads are categorized into different groups for primary frequency control with specific time-frequency control for each group of loads. During the unplanned islanding of the renewable energy sources, in case of disturbances, system frequency decreases gradually. To guarantee the frequency stability of the renewable energy system due to unplanned islanding, a control strategy of distributed generations (DGs), loads, and energy resources is presented in \cite{borsche2016stochastic}. 

Integration of RESs degrades the power quality at the point of common coupling (PCC) and  injects harmonic components, that must not exceeds specific limits,  to electrical networks \cite{farhoodnea2013power}. Power loss in the circuit and  communication system interference are two major problems due to high-level harmonics injection by the PE converters of PV and wind generation systems\cite{singh2007improved,asiminoael2007detection,an2016overview}.  It is imperative to improve the power quality by adopting several measures to ensure smooth system operation \cite{hossain2018analysis}.  

According to the aforementioned issues and their importances, this paper provides a broad view of the several challenges and opportunities in highly renewable integrated systems. Several challenges, such as total inertia reduction, low fault ride through capability, high uncertainties, voltage and frequency fluctuation, and low power quality, are well documented in this review article. In addition, several methodologies are also discussed to solve each of the above-mentioned problems. Some gaps are clearly mentioned in the current studies which could be filled by cutting edge technologies as form of new contribution from the researches and industry personnel.      
  
The paper is organized as follows: Section 2 provides frequency instability issues of RESs integrated system and possible solution methodologies; fault ride through and stability issues are addressed in section 3; power quality issues in RESs integration and several solution techniques are discussed in section  4;   modeling of uncertainty and optimization techniques in uncertainty reduction are discussed in section 5; current challenges for RESs integration and some future works are recommended in section 6; and finally, section 7 summarizes the major conclusions of this review.

\section{Low inertia and frequency issues} \label{SecIntertia}
Integration of renewable energy resources (RESs), both solar photovoltaic (PV) and wind, reduces the total inertia of the system due to the replacement of classical synchronous generators \cite{morren2006wind}. Although the variable speed wind turbines have inertia, it is effectively decoupled from the system, thus, it can not assist improving frequency response, due to connection of wind turbine to the network through the power electronic converters. Moreover, solar PV plants can not provide any inertia to the power system, which further degrades the frequency response. 
Therefore, high-level penetration of RESs to the system with the replacement of classical synchronous generator reduces overall inertia and increases rate of change of frequency (ROCOF) which activates load-shedding controller, even at small load-generation mismatch \cite{fini2019frequency}.
Furthermore, reduction in reserve power, due to replacement of reserve generating units, causes frequency deviation \cite{ulbig2014impact}. Thus, it is imperative to design new controllers for RESs to emulate the behavior of synchronous generator in order to improve frequency response of the system. The following subsections describe different control techniques of solar PV and wind systems to improve the frequency response for stable operation.

\subsection{Wind based system}
\begin{figure}[th]
\begin{center}
\includegraphics[scale=0.68]{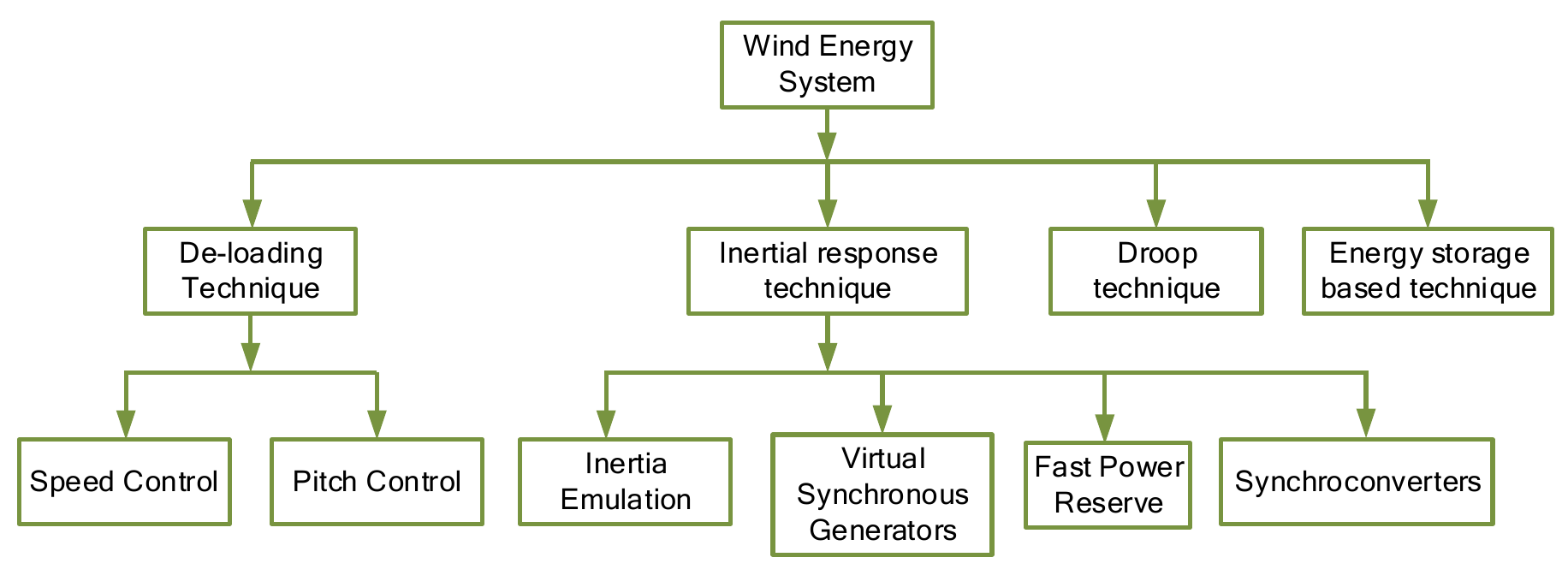}
\caption{Inertia and frequency support techniques by wind system}
\label{fig:classification}
\end{center}
\end{figure}

The dynamic behavior of the power system is studied with the swing equation of machine as below \cite{tamrakar2017virtual}.

\begin{equation} \label{genload}
P_{mech}-P_{elet}=J\omega\frac{d\omega}{dt}
\end{equation} 
where, $P_{mech}$ is the mechanical power input to the machine, $P_{elet}$ is the electrical power, $J$ is the moment of inertia, $\omega$ is the system frequency in rad/sec. The strength of the power system, either weak or strong, can be understood with the total inertia of the system as represented by the following equation. 
\begin{equation} \label{H}
H=\frac{J\omega}{2S}
\end{equation} 
where, $S$ is the summation of the apparent power of all generators. Now, simplifying equation (\ref{genload}) and (\ref{H}), following equations are obtained.

\begin{equation} \label{equ3}
\frac{2H}{\omega}\frac{d\omega}{dt}=\frac{P_{mech}-P_{elet}}{S}
\end{equation} 
\begin{equation} \label{equ3}
\frac{2H}{f}\frac{df}{dt}=\frac{P_{mech}-P_{elet}}{S}
\end{equation}
where, $f$ is the system frequency in Hz and $\frac{df}{dt}$ is the rate of change of frequency (ROCOF). Thus, ROCOF is inversely proportional to the system inertia. In order to meet the stability criteria, the system requires additional inertia while integrating more RESs. The concept of virtual inertia technologies, using power electronic converters, energy storage systems (ESSs), and control algorithms, which release the stored kinetic energy of wind turbine, are presented in the literature to support the frequency of RESs integrated system. The wind energy system can help in stabilizing grid frequency with different techniques such as de-loading technique, inertial response technique, and droop technique. Each of these techniques can be sub-categorized into different techniques \cite{fini2018determining}. Classification of inertia and frequency support techniques by wind turbine is presented in Figure \ref{fig:classification}.   

\subsubsection{De-loading technique} \label{deloads} 
\begin{figure}
        \centering
        \begin{subfigure}[h]{0.47\textwidth}
                \includegraphics[width=\textwidth]{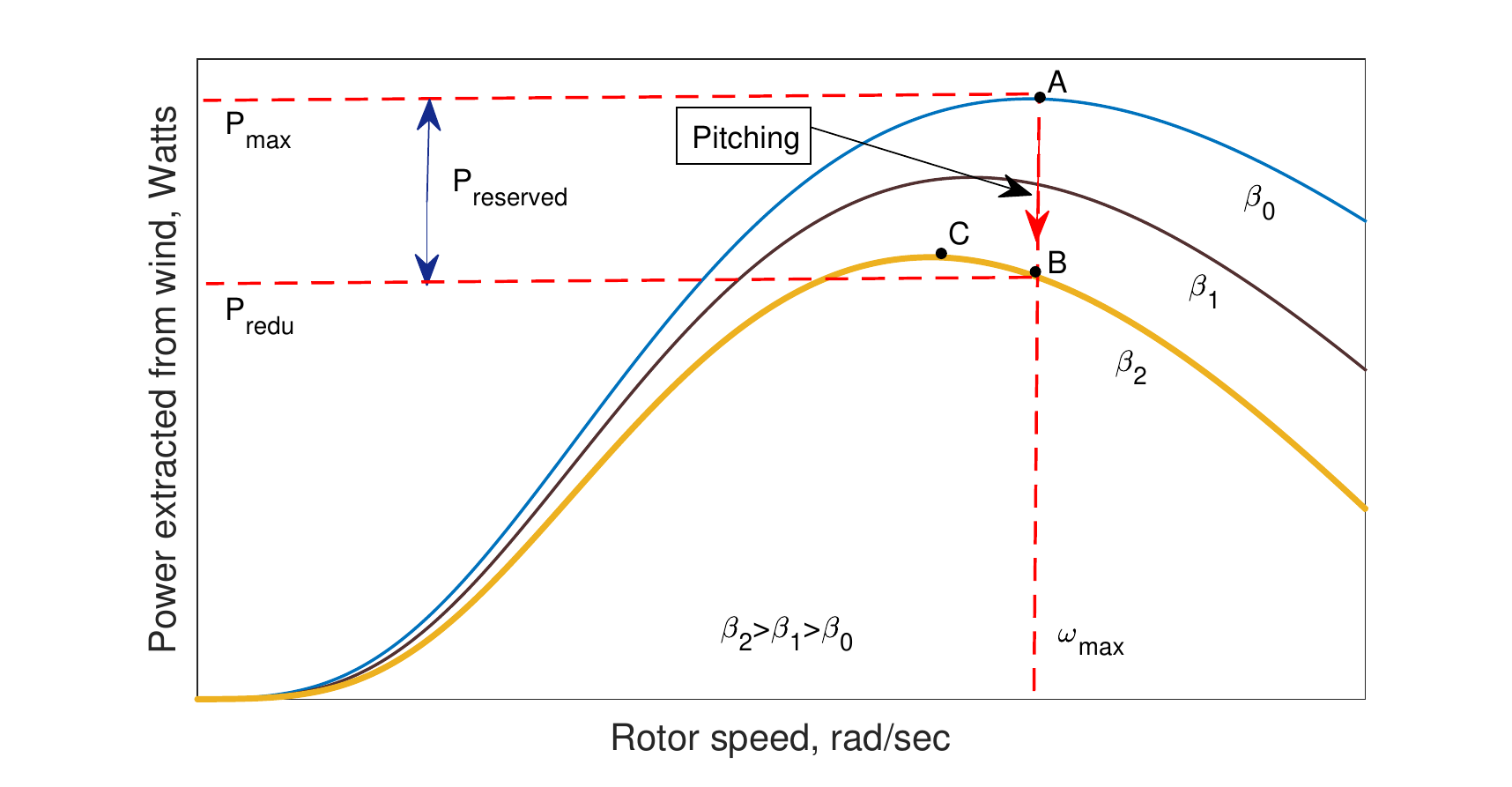} 
                \caption{ }
                \label{fig:wind1}
        \end{subfigure}
        \quad%
        ~ 
         \begin{subfigure}[h]{0.47\textwidth}
                \includegraphics[width=\textwidth]{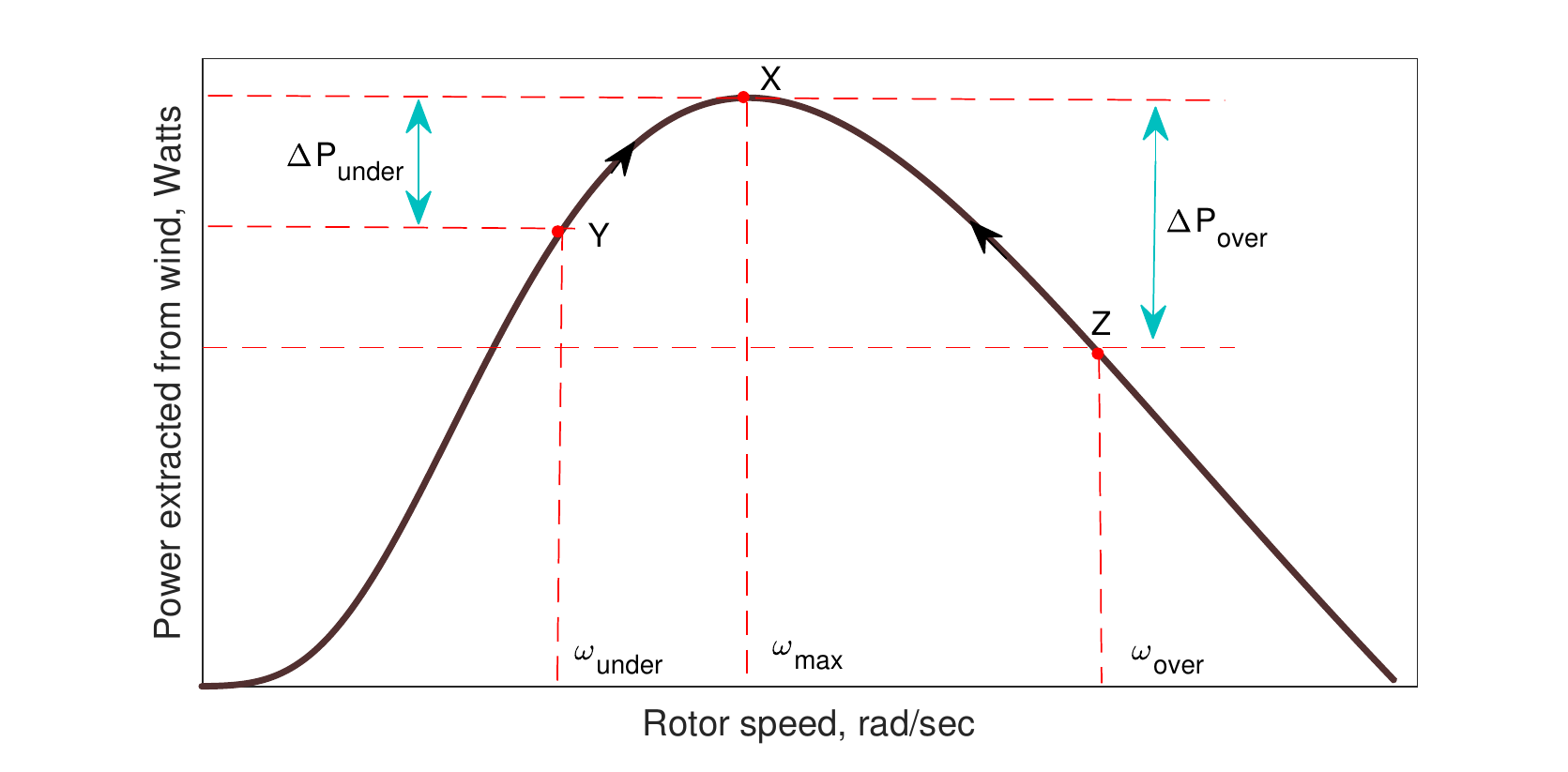}
                \caption{ }
                \label{fig:wind2}
        \end{subfigure}
       \vspace{-10pt}
                  \caption{De-loading techniques for wind turbine by (a) pitch control (b) speed control}
        \label{fig: deloadwind}
\end{figure}

De-loading technique, ability of the wind system to provide reserve power, is evolved to address the frequency deviation \cite{pradhan2015enhancement}. This technique shifts the optimal operating point of wind turbine to the reduced power level point, as a result wind system provides some reserve generation, which can participate in frequency regulation \cite{ma2010working}. De-loading technique mainly consists of two operation modes: pitch-angle control mode and speed control mode. The former one shifts the pitch angle from zero to some higher values and the latter one shifts the turbine speed left or right from the maximum power point, as shown in Figure \ref{fig: deloadwind}. 

As shown in Figure \ref{fig:wind1}, pitch control is achieved by increasing pitch angle from some lower value, $\beta_0$, to some higher value, $\beta_2$, for a constant wind speed, corresponding to rotor speed at maximum power point, $\omega_{max}$. In this way, operating point shifts from point A to B instead of C, which can provide reserve power $P_{reserved}$, during frequency deviation of the system \cite{vzertek2012optimised,fini2018determining,ma2010working}.

Speed control mode de-loading technique has two possibilities, such as over-speed control and under-speed control. In former one, rotor speed controller adjusts the rotor  speed at somewhat higher value, for example, $\omega_{over}$, for a constant pitch angle. In case of wind system to participate in frequency regulation, rotor speed is adjusted back to the point corresponding to maximum power point, $\omega_{max}$. Thus, shifting the operating point from Z to X provides additional reserve power, $\Delta$ $P_{over}$, during frequency instability due to generation and load mismatch.
 In latter one, see Figure \ref{fig:wind2}, rotor speed is controlled to somewhat lower value, $\omega_{under}$, which is below the maximum power point speed, $\omega_{max}$. Thus, operation of wind generator at this point has a reserve power, $\Delta$ $P_{over}$. However, over-speed control mode is preferable than under-speed control mode, since in latter one, speed is increased from $\omega_{under}$ to $\omega_{max}$ utilizing some power extracted from the turbine. This shows opposite behavior during the first interval of the frequency response and is considered as 'detrimental strategy' 
\cite{janssens2007active,ramtharan2007frequency}.

The de-loading techniques provide reserve power to support the frequency of the system. However, the amount of power is not specific. In \cite{wu2012coordinated,zhangjie2012control}, a combined pitch angle and over-speed controller is proposed to participate in frequency regulation based on the request from the system operator for specific amount of power. Further improvement in frequency control is achieved with the pitch and over-speed control, coordinated with droop control \cite{zhangjie2012control}. This control topology is tested for doubly fed induction generator (DFIG) based wind system. Most of the deloading techniques involve operation of wind turbine in deloaded mode for long-term, which is responsible for economic loss for the wind turbine owners. To minimize this economic loss, a coordinated strategy is presented in \cite{peng2019coordinated}. In this strategy, DFIG does not need to operate in deloaded mode for long-term, instead, it can operate in maximum power point tracking (MPPT) mode while there is no need for frequency support of the system.

\subsubsection{Inertial response technique}
Conventional synchronous generator can release the kinetic energy, which is stored in the rotating mass, automatically to the grid; however, renewable energy resources (RESs) can not do the same due to the decoupling between rotating mass and grid through the power electronic converters. To resolve this issue, inertial response techniques are evolved, which can be categorized into different ones such as inertia emulation, virtual synchronous generator, fast power reserve, synchroconverter. The main idea behind these techniques is to emulate the behavior of classical synchronous generators through the RESs. 

In inertia emulation technique, the kinetic energy stored in the rotating mass is released with the new control loops \cite{kerdphol2017virtual}. Generally, one-loop and two-loop control strategies are used for inertia emulation by the wind systems. In one-loop control strategy, kinetic energy stored in the rotating blades is released based on rate of change of frequency (ROCOF) with a single loop only, while the latter one does the similar task with two loops for better frequency response.

   \begin{figure}
        \centering
        \begin{subfigure}[h]{0.44\textwidth}
                \includegraphics[scale=0.88,width=\textwidth]{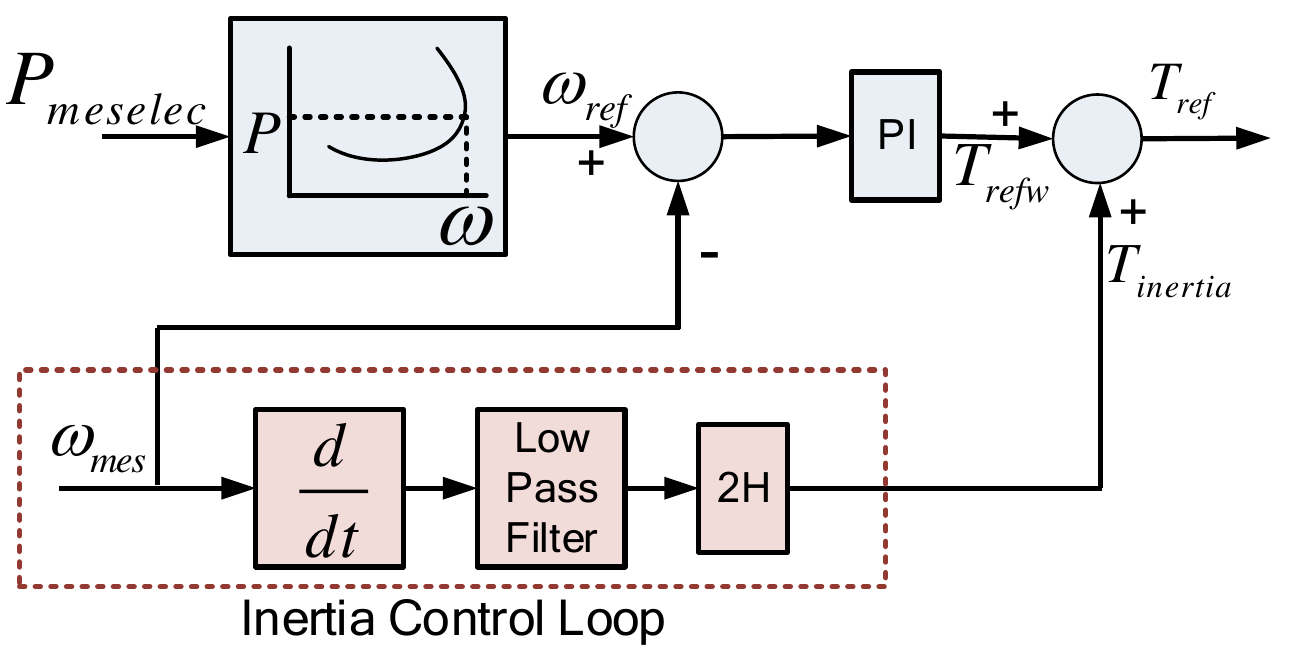} 
                \caption{ }
                \label{fig:inertia1}
        \end{subfigure}
        \quad%
        ~ 
         \begin{subfigure}[h]{0.44\textwidth}
                \includegraphics[scale=0.88,width=\textwidth]{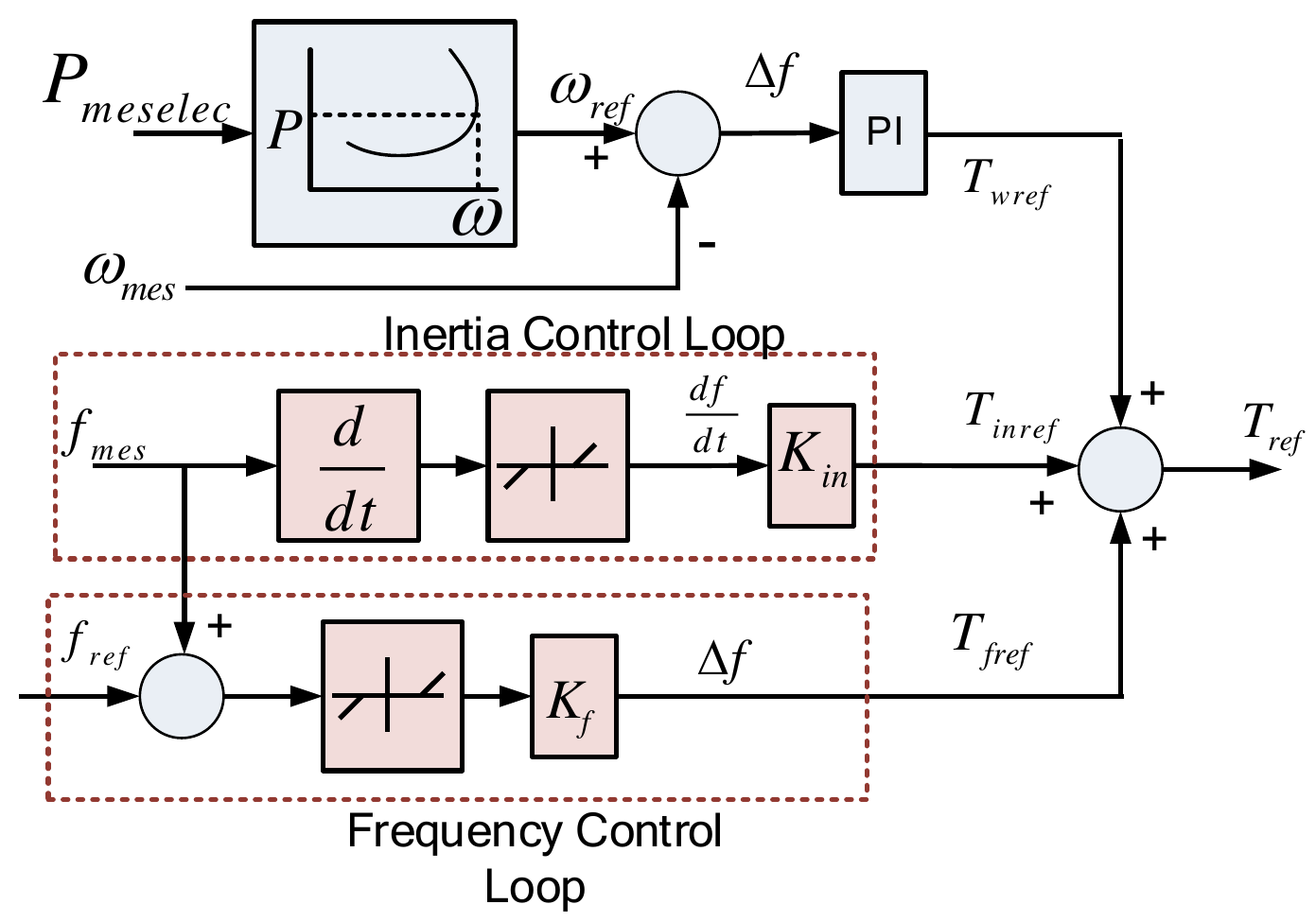} 
                \caption{ }
                \label{fig:inertia2}
        \end{subfigure}
       \vspace{-10pt}
                  \caption{Inertia emulation techniques (a) one-loop control (b) two-loop control}
        \label{fig: inertia}
\end{figure} 
 
As shown in Figure \ref{fig:inertia1},  MPPT controller loop determines the reference speed of the rotor, $\omega_{ref}$, which is processed by the PI controller to provide torque reference, $T_{refw}$, corresponding to maximum power in normal operating condition. However, during frequency deviation, inertia control loop is enabled to provide additional torque, $T_{inertia}$. Due to this additional torque, generator speed is slowed down, and the kinetic energy stored in the rotor is released \cite{morren2006inertial,ekanayake2004comparison}. However, the main disadvantage of this method is that the amount torque provided by the additional control loop is constant, which is responsible for rapid reduction of rotor speed as well as delay in controller operation. To overcome this issue, in \cite{wu2013towards}, an inertia response technique is presented to dynamically adapt the inertia constant during frequency response support with an idea to increase the inertia constant as long as the frequency of the system continues to decline. This strategy is applied for DFIG based system in \cite{zhang2013comparison} and compared for different values of $K_{in}$ and $K_f$ as shown in Figure \ref{fig:inertia2}. 

\begin{figure}[h!]
\begin{center}
\includegraphics[scale=0.58]{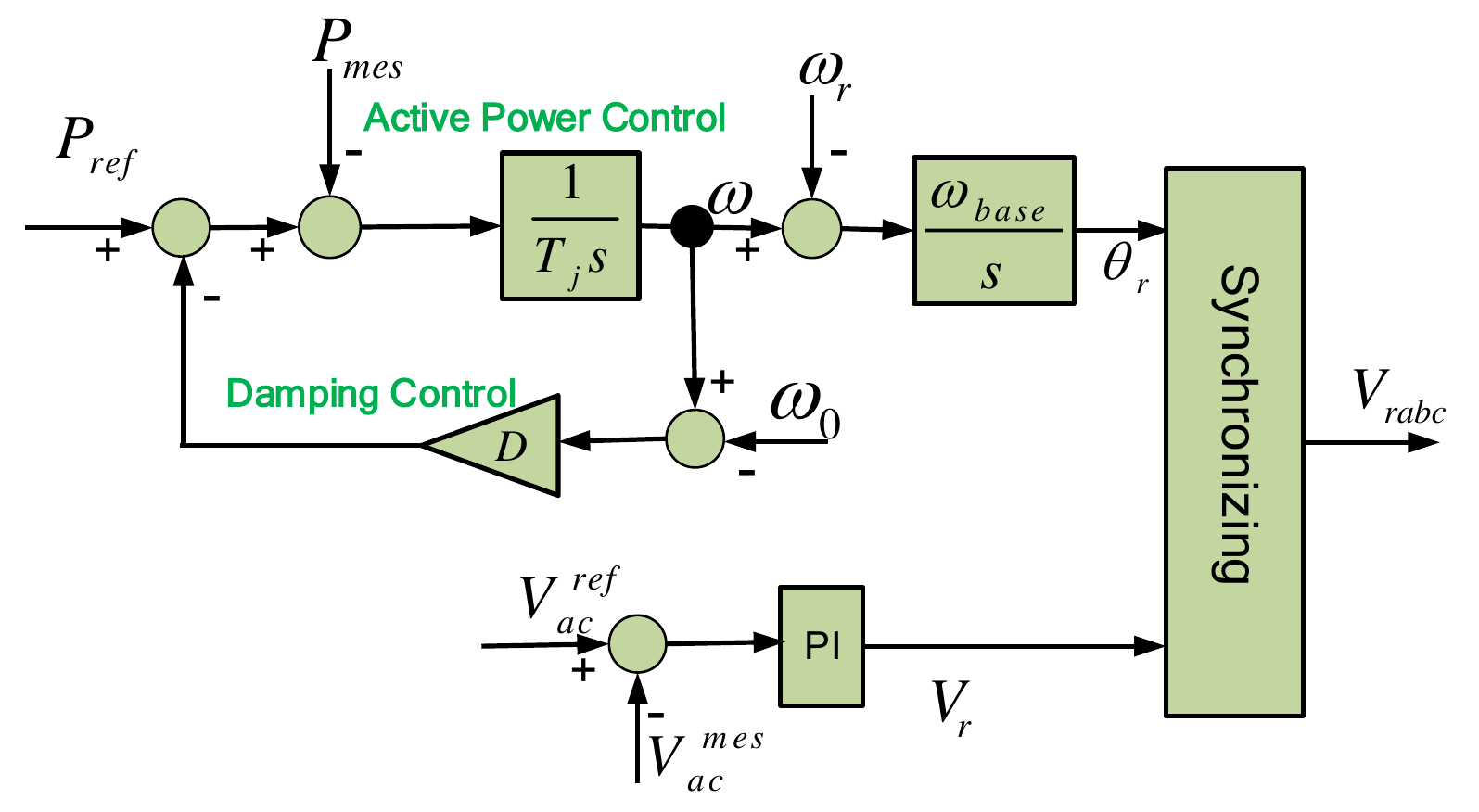}
\caption{Synchroconverter control for DFIG based wind system for inertia support}
\label{fig:syn_con}
\end{center}
\end{figure}

Most of the inertial response techniques adopt modification of vector control, which is based on classical phase locked loop (PLL) and voltage source converter (VSC). For example, in \cite{ekanayake2004comparison,hughes2005control,goksu2010review}, some techniques are presented to allow the wind turbines to emulate inertia by providing additional signals based on ROCOF. However, the classical synchronizing device, PLL, may have some negative impacts on system stability, which is reported in the literature \cite{jovcic2003vsc,midtsund2010evaluation}. To resolve this issue, another inertial response technique is developed which is called synchroconverter. The synchroconverter topology, which adopts the synchronous generator model based topology, is evolved to support the frequency of weak grid by the wind generation system. The synchroconverter provides a enhanced PLL or a sinusoid-locked loop, which makes wind system inherently capable to maintain synchronism through the active power control \cite{zhong2012control,wang2015virtual}. Detailed control strategy of synchroconverter is shown in Figure \ref{fig:syn_con}.    

Since there is no damper winding in DFIG and the resistances in rotor and stator windings are low, it requires additional damping controller as depicted in Figure \ref{fig:syn_con}. As shown in Figure \ref{fig:syn_con}, the DFIG based wind system is synchronized with the grid based on active power control scheme without relying on PLL. The error between measured and reference active power causes the regulation of $\omega$ via the virtual inertia. Phase angle i.e. the rotor voltage angle, for synchronization, is obtained by direct integration of slip frequency, $\omega_{slip}$. 

\begin{figure}[H]
\begin{center}
\includegraphics[scale=0.65]{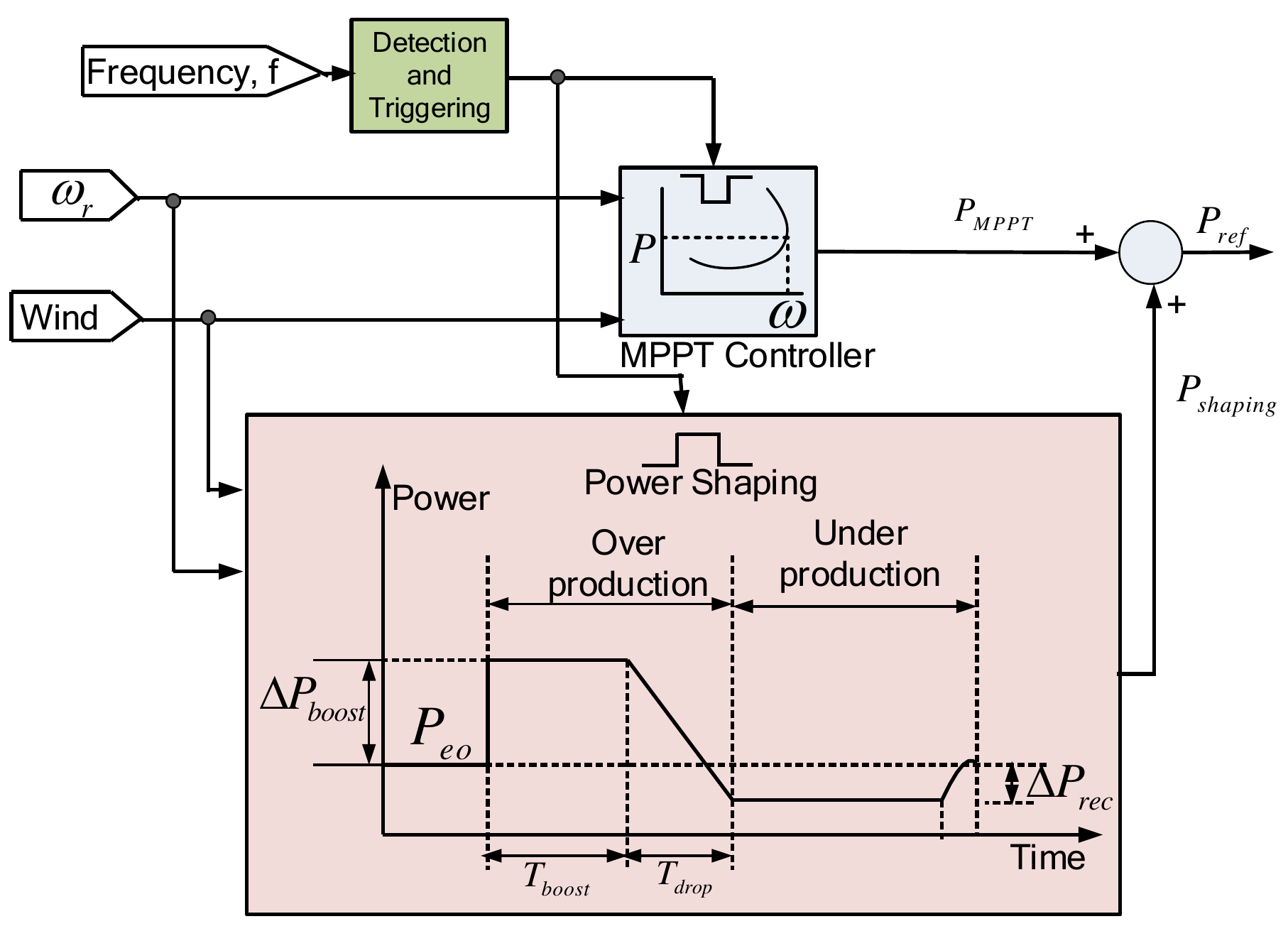}
\caption{Fast power reserve technique for frequency support}
\label{fig:pow_res}
\end{center}
\end{figure}

Another method for inertial response is fast power reserve technique, which is based on supplying the stored kinetic energy of the rotating turbine to the grid by means of some modified control strategies \cite{ullah2008temporary,hansen2014analysis,el2011short}. However, detailed control strategies are not documented in references \cite{ullah2008temporary,hansen2014analysis}. In fast power reserve technique, frequency deviation is used as input to the detection and triggering circuit \cite{el2011short}. In normal system operation, detection and triggering circuits enable the MPPT control loop and bypass the power shaping loop. However, in abnormal condition, when the system frequency deviation is more than the threshold level, power shaping loop is enabled and MPPT loop is disabled as shown in Figure \ref{fig:pow_res}. At this level, the wind generation system enters overproduction mode until the kinetic energy of the wind turbine is completely discharged. Afterward, wind generation system returns to MPPT mode. It is worth mentioning that, transition between overproduction to MPPT model may lead to underproduction phase, in which power is reversed from grid to turbine. To avoid this unexpected operation, instead of sharp transition, a sloped transition is presented as shown in Figure \ref{fig:pow_res}. 

\subsubsection{Droop control technique}

\begin{figure}[H]
\begin{center}
\includegraphics[scale=0.68]{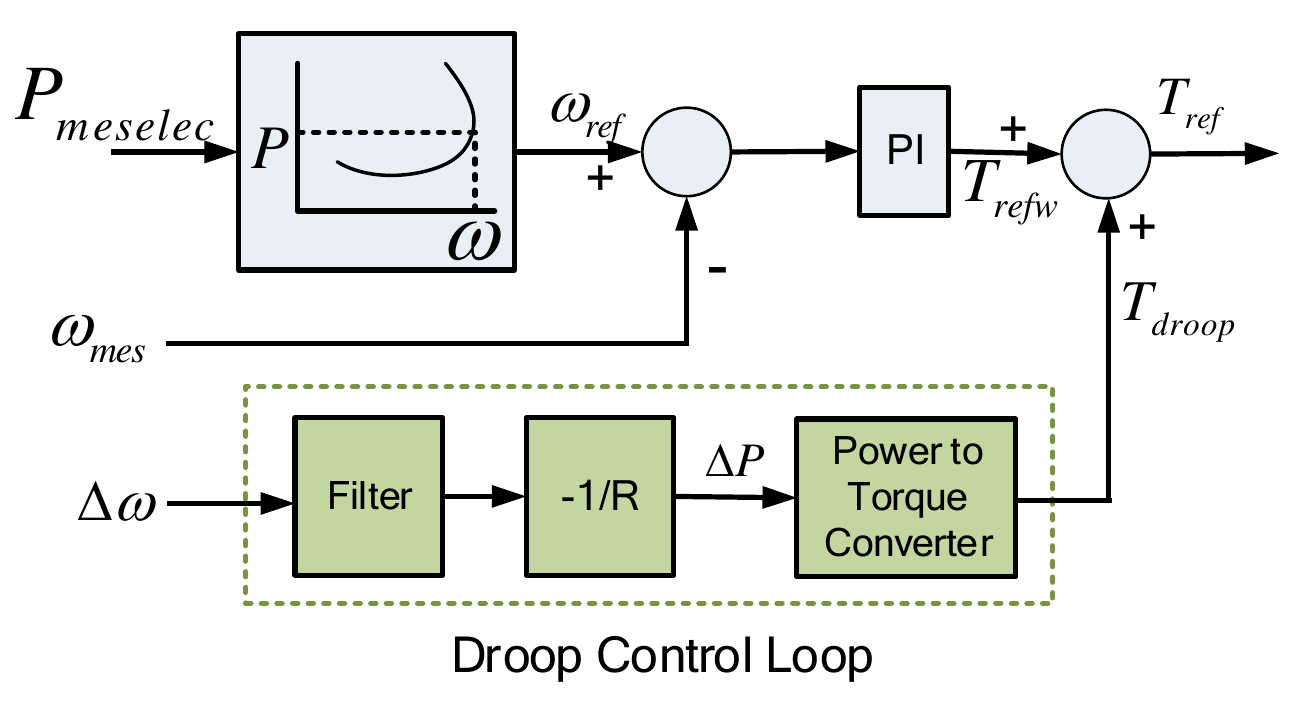}
\caption{Droop control technique for wind system}
\label{fig:f6_droop}
\end{center}
\end{figure}

The droop controller, which adjust the output power with the variation of system frequency based on the droop setting, is evolved to support primary frequency control \cite{jahan2019primary}. The droop gain and power relation is given by the following equation. 
\begin{equation} \label{equ5}
\Delta P_{droop}=K_{droop}\Delta f
\end{equation} 
In this work, the droop controller is only enabled when the system frequency deviation exceeds a specific limit ($|\Delta f|>|a|=0.075$). The droop gain ($\Delta K_{droop}$) is optimized by perturb and observe method. Sufficient power is kept as reserved by deloading technique, as discussed in section \ref{deloads}, which is then used by the centralized droop controller in response to the frequency deviation.  
Similar to conventional synchronous generator equipped with speed governor for frequency regulation, wind energy system can support frequency by adjusting active power according to droop setting \cite{yao2011control,josephine2014estimating}. Active power is adjusted with the frequency deviation given by the following equation.
\begin{equation} \label{equ6}
\Delta P=P_{low}-P_{high}=-\frac{\omega_{high}-\omega_{low}}{R}
\end{equation} 
where, $R$ is the droop constant, $P_{low}$ is the low power, $P_{high}$ is the high power, $\omega_{low}$ is the low frequency, $\omega_{high}$ is the high frequency. Detail droop control structure is shown in Figure \ref{fig:f6_droop}.

As shown in Figure \ref{fig:f6_droop}, the reference torque is generated as a sum of MPPT controller loop torque and droop control loop torque. Depending on the frequency deviation of the system, torque command is modified by the droop controller to stabilize system frequency.

\subsubsection{Energy storage based technique} 
Modification of the existing control strategies as well as introduction of several new strategies are presented in the literature  to solve low inertia and frequency issues of variable speed wind system as discussed in the previous sections. However, it is worth mentioning that the strategies discussed  earlier have low reliability issue due to varying nature of wind. A potential solution to this problem is to integrate wind energy to the grid through energy storage systems (ESSs), such as battery, superconducting magnetic energy storage (SMES), fly wheel energy storage, super capacitors \cite{worku2016power,miao2015coordinated,ye2017efficient,miao2015coordinated,zhang2015fuzzy}. Frequency response is improved for wind energy system  using battery energy storage system (BESS) in \cite{khalid2012optimal} and a combined BESS and automatic generation control (AGC) strategy is presented in \cite{zhao2015review} for improved frequency response. Some of the ESSs has higher energy density whereas some of them has higher power density. Thus, in \cite{esmaili2013hybrid,mendis2013management}, hybrid ESS based frequency support technique is presented to harness both high power and energy densities.

\begin{figure}[h]
\begin{center}
\includegraphics[scale=0.68]{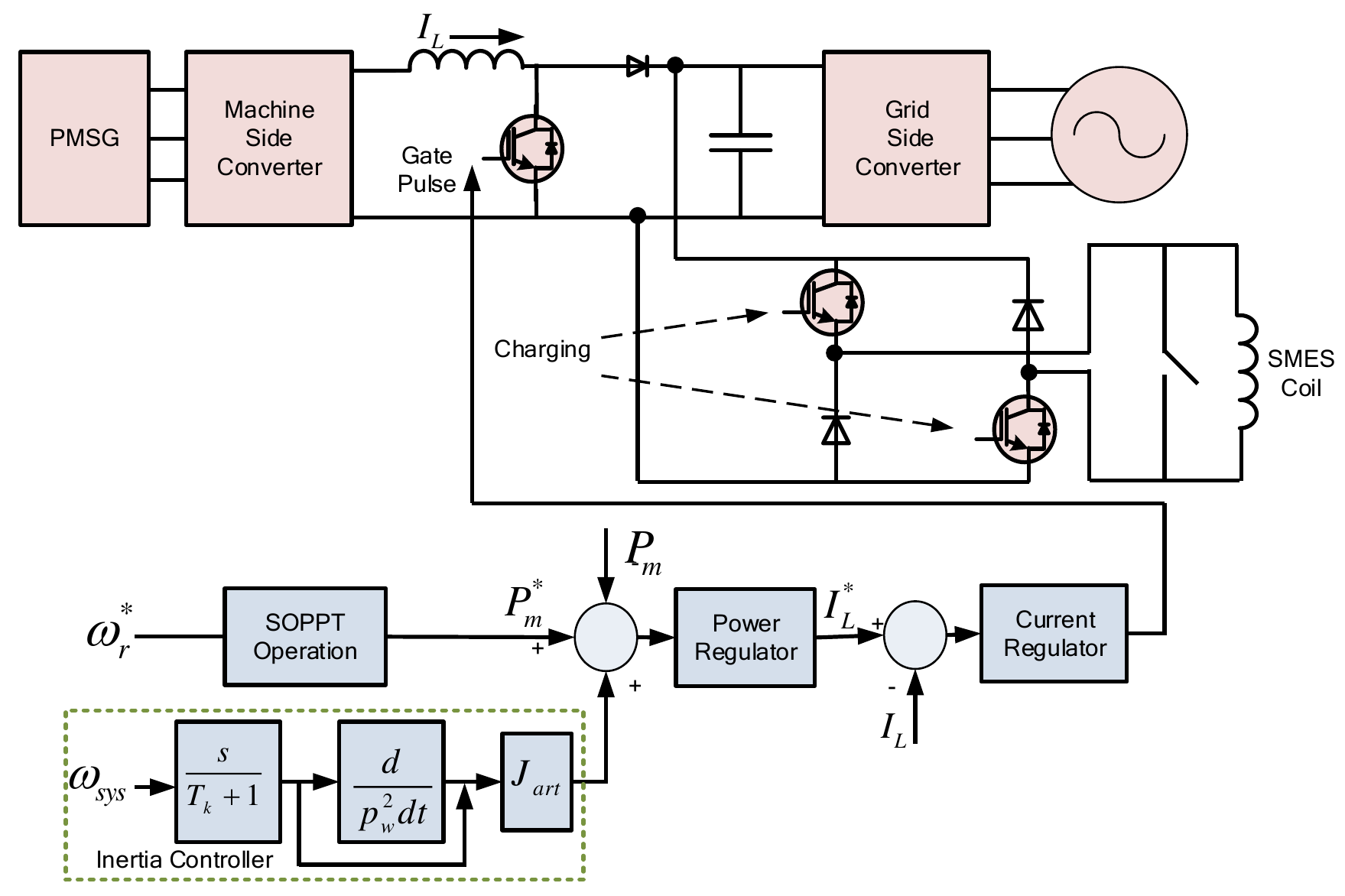}
\caption{Energy storage based frequency support for PMSG wind system}
\label{fig:ess_wind1}
\end{center}
\end{figure}
\begin{figure}[h]
\begin{center}
\includegraphics[scale=0.68]{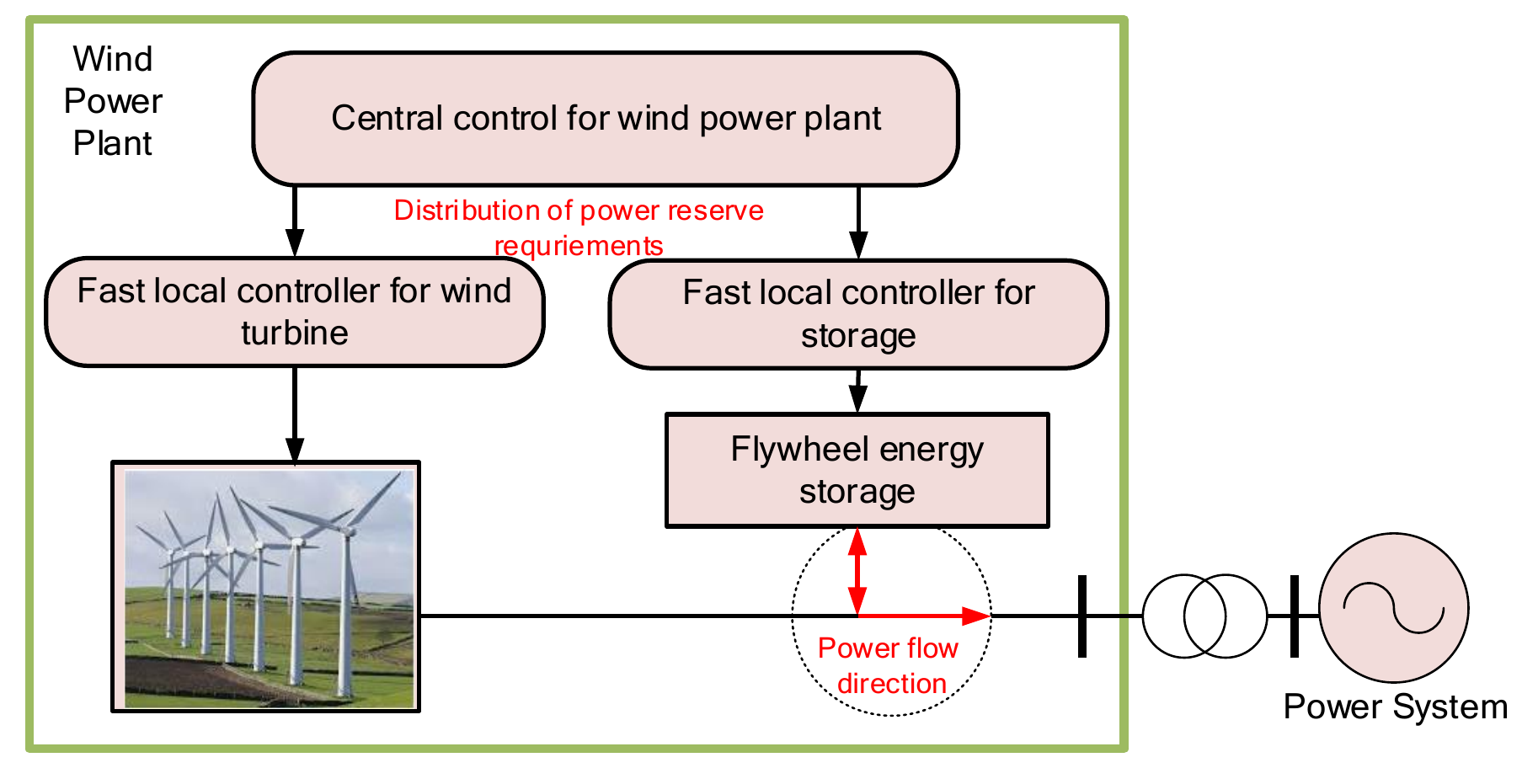}
\caption{Frequency regulation by wind farm and flywheel energy storage}
\label{fig:ess_wind2}
\end{center}
\end{figure}

Since the natural inertia of variable speed wind generator is much lower than classical synchronous generator, SMES, which is fast responding compared to other energy storage device, is presented to improve primary frequency response of permanent magnet synchronous generator (PMSG) system  \cite{musarrat2018enhanced}. The improvement in frequency response of PMSG wind system is achieved with artificial inertial controller which controls the boost controller in the DC link as shown in Figure \ref{fig:ess_wind1}. In stable operation, the difference between electromagnetic torque and mechanical torque is zero; however, in worst case scenario, the system frequency deviates, and change in the reference torque reduces the rotor speed to emulate the inertia. The inertia controller provides duty to the boost converter, which adjusts the power output and torque for primary frequency support by controlling the current through the reactor, $I_L$. DFIG based wind system frequency response improvement is presented in\cite{diaz2015coordinated} with battery and flywheel energy storage. In \cite{diaz2015coordinated}, flywheel based storage is considered as an integral part of wind power plant to provide reserve power indicated by the system operator for primary frequency control as shown in Figure \ref{fig:ess_wind2}. Total power reserve required by the system is distributed between the wind turbine and flywheel energy storage. Power reserves of wind turbine and energy storage are activated by the local controls immediately after the frequency deviation exceeds a predefined value. The central control is employed to supervise the activation of local controls.    

\subsection{Solar based system}
\begin{figure}[h]
\begin{center}
\includegraphics[scale=0.68]{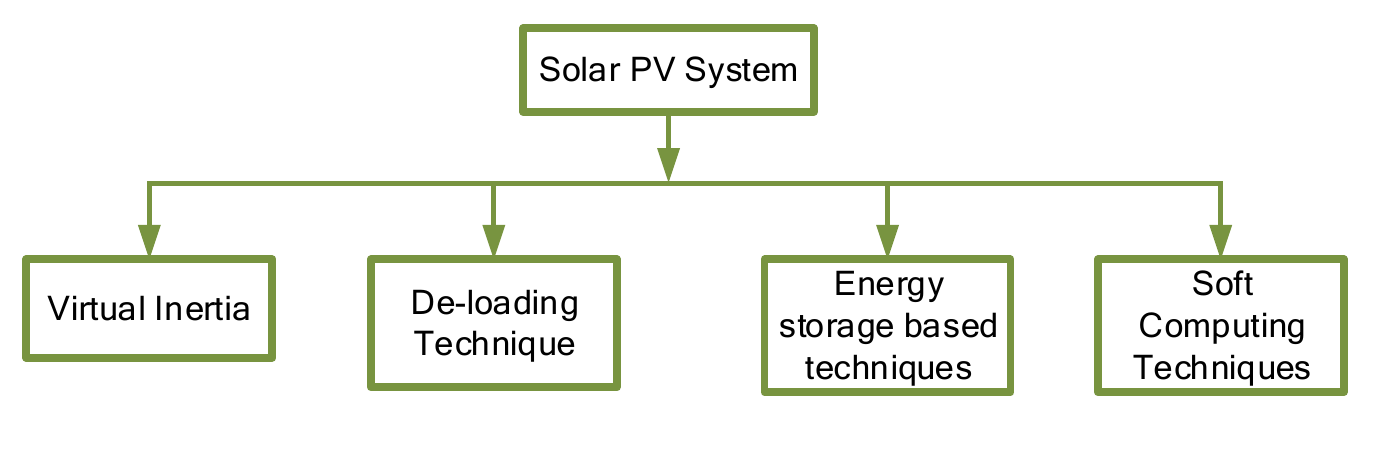}
\caption{Frequency and inertia support by PV system}
\label{fig:solar_classifi}
\end{center}
\end{figure}

Grid connected solar photovoltaic (PV) system can participate in frequency regulation during positive frequency excursion, increase in system frequency due to higher generation than load, by reducing the output of PV. However, it can not participate in frequency regulation during negative frequency excursion, since it operates at maximum power point having no  reserve margin. For the PV system to participate in frequency regulation during negative frequency excursion, some reserve must be kept by deloading or some other techniques. Mainly, three possible techniques are presented in the literature: charging a energy storage devices \cite{sa2015intelligent,hill2012battery}, operating PV  system in reduced power output mode by deloading \cite{kakimoto2009power,xin2013new,zarina2014exploring,alatrash2012generator} and inertial response technique \cite{nanou2015generic}. Several techniques presented in the literature for frequency and inertial support from PV system is shown in Figure \ref{fig:solar_classifi}. 
 
\subsubsection{Inertial response technique}
\begin{figure}[h]
\begin{center}
\includegraphics[scale=0.68]{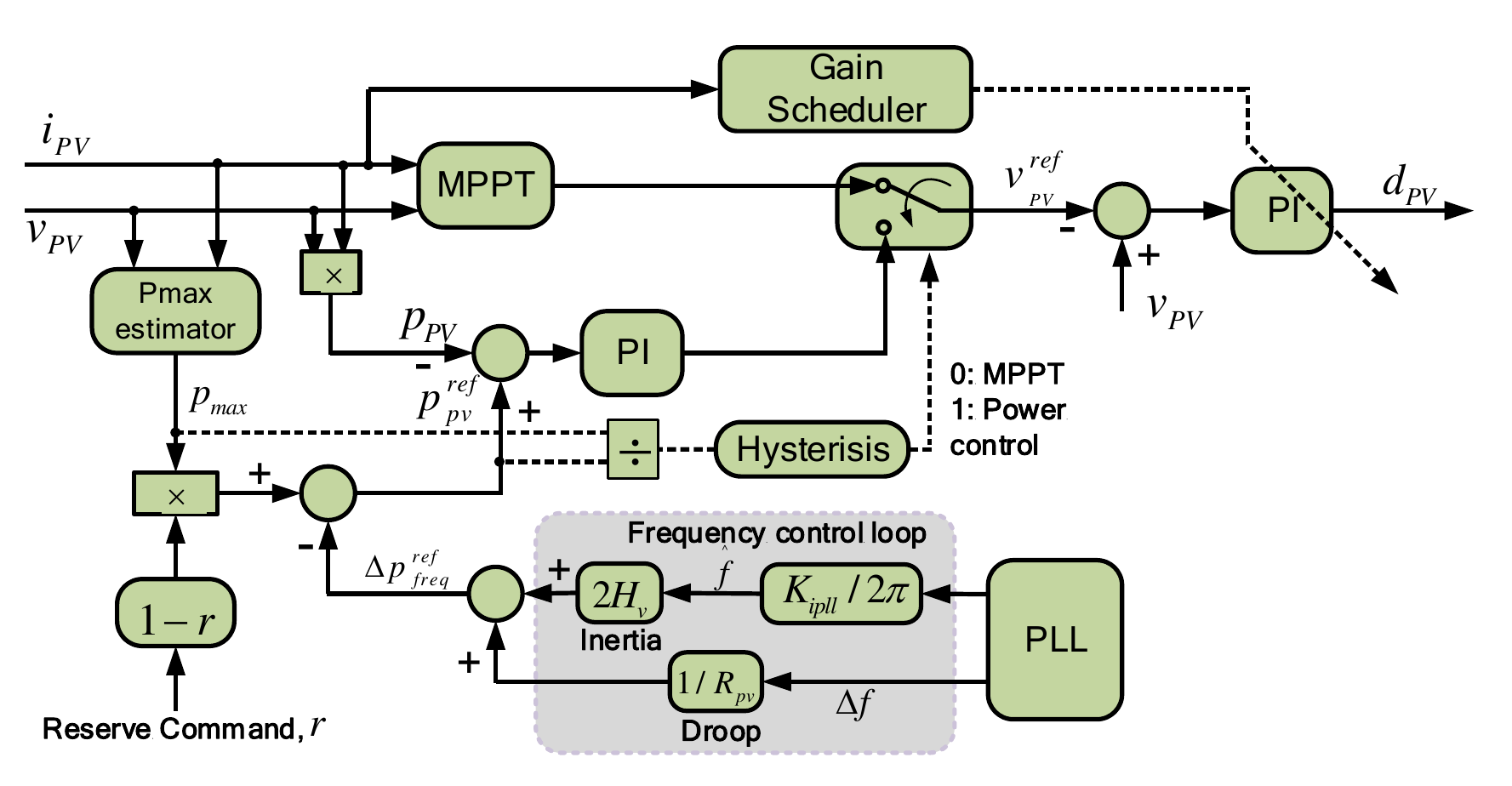}
\caption{Primary frequency regulation and inertia emulation by PV}
\label{fig:pv_inertia}
\end{center}
\end{figure}

Primary frequency control by PV system is presented in \cite{nanou2015generic} with inertia emulation technique. The inner and outer control loops are implemented to generate duty cycle for the DC/DC converter of PV system as shown in Figure \ref{fig:pv_inertia}. The former regulates the PV array voltage to its reference value while the latter regulates the PV power to the reference value either by MPPT controller or power controller. The reference power for the outer controller is given by the following equation.
\begin{equation} \label{pv_inr}
\ p_{pv}^{ref}=(1-r).p_{max}-\Delta p_{freq}^{ref}
\end{equation}  
where $r$ is the reserve power set by the system operator, $P_{max}$ is the estimate of maximum available power, $\Delta p_{freq}^{ref}$ is the output of frequency controller. 
The frequency controller, which comprises proportional and derivative terms, gives the following frequency dependent PV power reference. 
\begin{equation} \label{pv_inr}
\Delta p_{freq}^{ref}=\Delta f/R_{pv}+2H_{pv}\hat{f} 
\end{equation} 
where, $R_{pv}$ is the droop constant, $H_{pv}$ is the virtual inertia gain and $\hat{f}$ is the rate of change of frequency.  
 
\subsubsection{De-loading technique}
\begin{figure}[h]
\begin{center}
\includegraphics[scale=0.68]{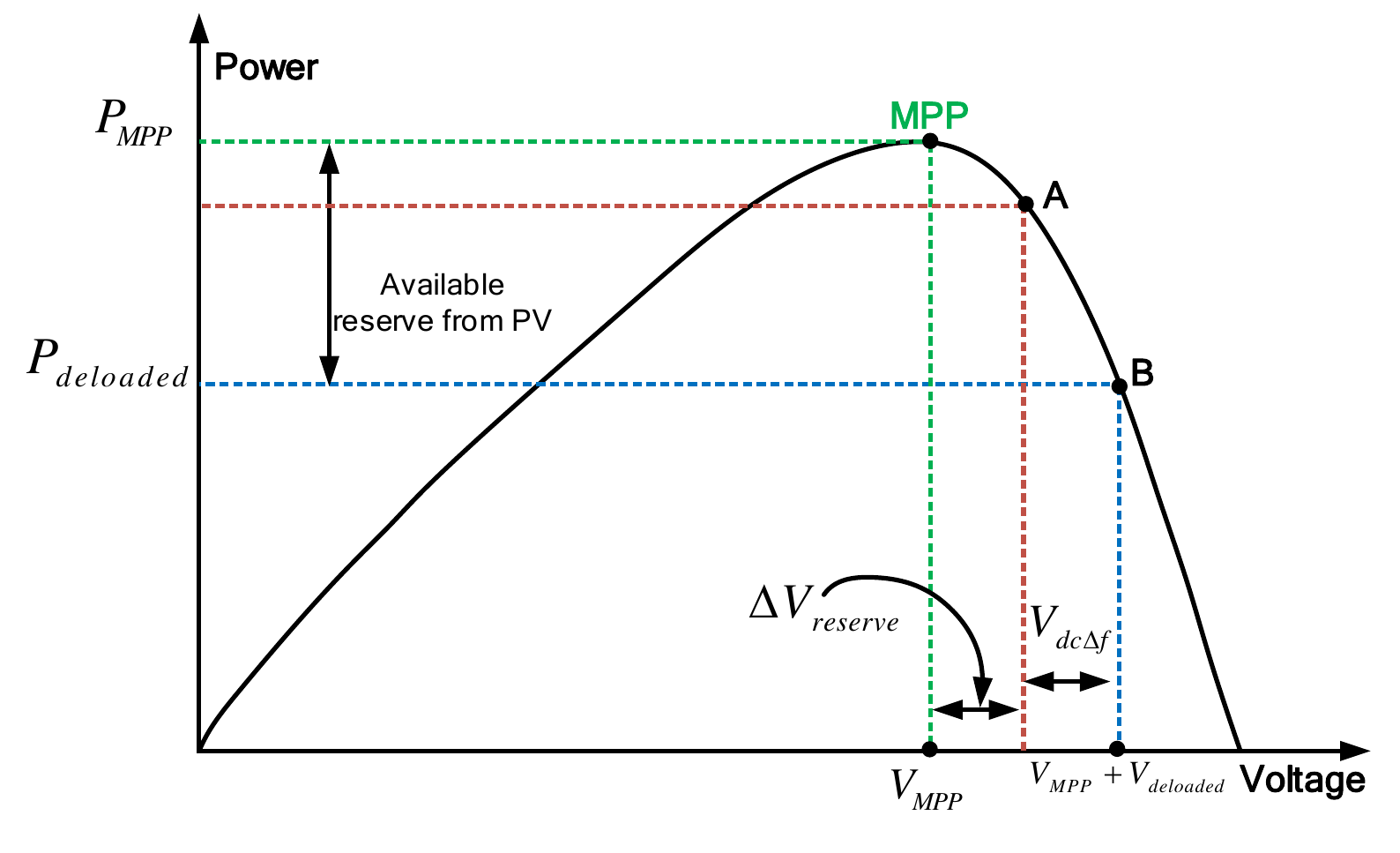}
\caption{PV system de-loading power-voltage curve}
\label{fig:pv_deload_curv}
\end{center}
\end{figure}

The PV system can provide reserve and support system frequency by delaoding technique which involves operation of PV system beyond the MPP as shown in Figure \ref{fig:pv_deload_curv}. The maximum power corresponds to point MPP with a voltage of $V_{MPP}$. As shown, instead of operating at MPP, the PV system operates at point $B$ having a total reserve of $P_{max}-P_{delaoded}$.   

\begin{figure}[h]
\begin{center}
\includegraphics[scale=0.68]{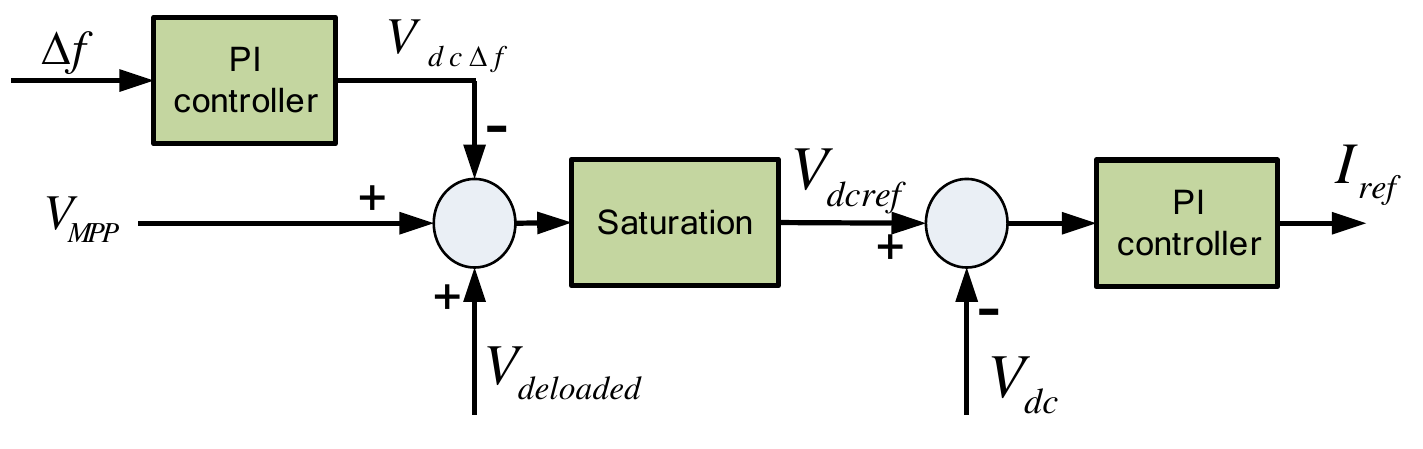}
\caption{PV system deloading controller}
\label{fig:pv_deload_controller}
\end{center}
\end{figure}

The deloading technique as presented in \cite{zarina2016power} is shown in Figure \ref{fig:pv_deload_controller}. As shown, the PV output power depends on both $V_{MPP}$ and system frequency deviation $\Delta f$ which is given by the following equation.
 \begin{equation} \label{pv_deload}
V_{dcref}=V_{MPP}+V_{deloaded}-V_{dc\Delta f}
\end{equation} 

However, the controller presented in Figure \ref{fig:pv_deload_controller} has non-uniform distribution of frequency regulation. This mainly happens for same amount of power release from the PV units having different reserve level. So, the PV units with less reserve reach MPP faster than other PV units with higher reserve and further frequency regulation can not be achieved by these units due to their operation at MPP. In order to address this issue, a modified controller is presented in \cite{zarina2014exploring}, in which, the output power delivered from each unit depends on the reserve, instead of delivering same amount of power from each unit. The modified reference voltage for this new controller is given by the following equation.    
\begin{equation} \label{pv_deload}
V_{dcref}=V_{MPP}+V_{deloaded}-V_{dc\Delta f}-(\Delta f* \Delta V_{reserve} *K_p)
\end{equation} 
where, $\Delta f$ is the system frequency deviation, $V_{reserve}$ is the voltage corresponding to reserve power, and $K_p$ is the gain of the proportional controller. 

Another deloading technique is is presented in \cite{yan2019novel} which is named as  adaptive de-loading technique  having three controller loops: droop controller, active power-voltage matching controller, vector controller. This technique provides a reserve power in PV system with a possibility to adjust quickly the output power of PV for frequency regulation of the grid.

\subsubsection{Energy storage based technique}

Energy storage devices can be used to mitigate the negative impact of high penetration of PV to the grid \cite{showers2019frequency,you2019energy} by reducing the active power variation. In \cite{you2019energy}, droop  and step response controllers with energy storage are presented to improve frequency response of two high PV system united states power grids. Performance of step response controller with energy-constrained high-power-density storage system is found slightly better than droop controller in this study. In \cite{wang2016design}, a battery energy storage system (BESS) is designed to support grid frequency, by BESS input current regulation, with an efficient DC-DC converter control. Additionally, the proposed controller is capable to improve fault ride through capability in case of different transients in the system.     

\subsubsection{Soft computing techniques} 

The output power fluctuation  of the PV system, as a change of the weather conditions, season, and
geographic location,  cause high-level frequency deviation of the power system. In \cite{ray2019robust,datta2010frequency}, soft commuting methods are applied in the PV system to reduce power fluctuation from the PV system for improving the frequency response. Depending on frequency deviation and average insolation of PV system, output power command is generated in \cite{datta2010frequency} using fuzzy logic controller. The method is capable to operate the PV system to near maximum power point which is better than the deloading technique. Another soft computing technique similar to the reference \cite{datta2010frequency} is presented in \cite{sakeen2013frequency} which combines fuzzy logic controller and particle swarm optimization  to generate output power command in order to improve frequency response of PV system. 

\section{Fault ride through (FRT) capability issues} \label{FRT}

\begin{figure}[h]
\begin{center}
\includegraphics[scale=0.68]{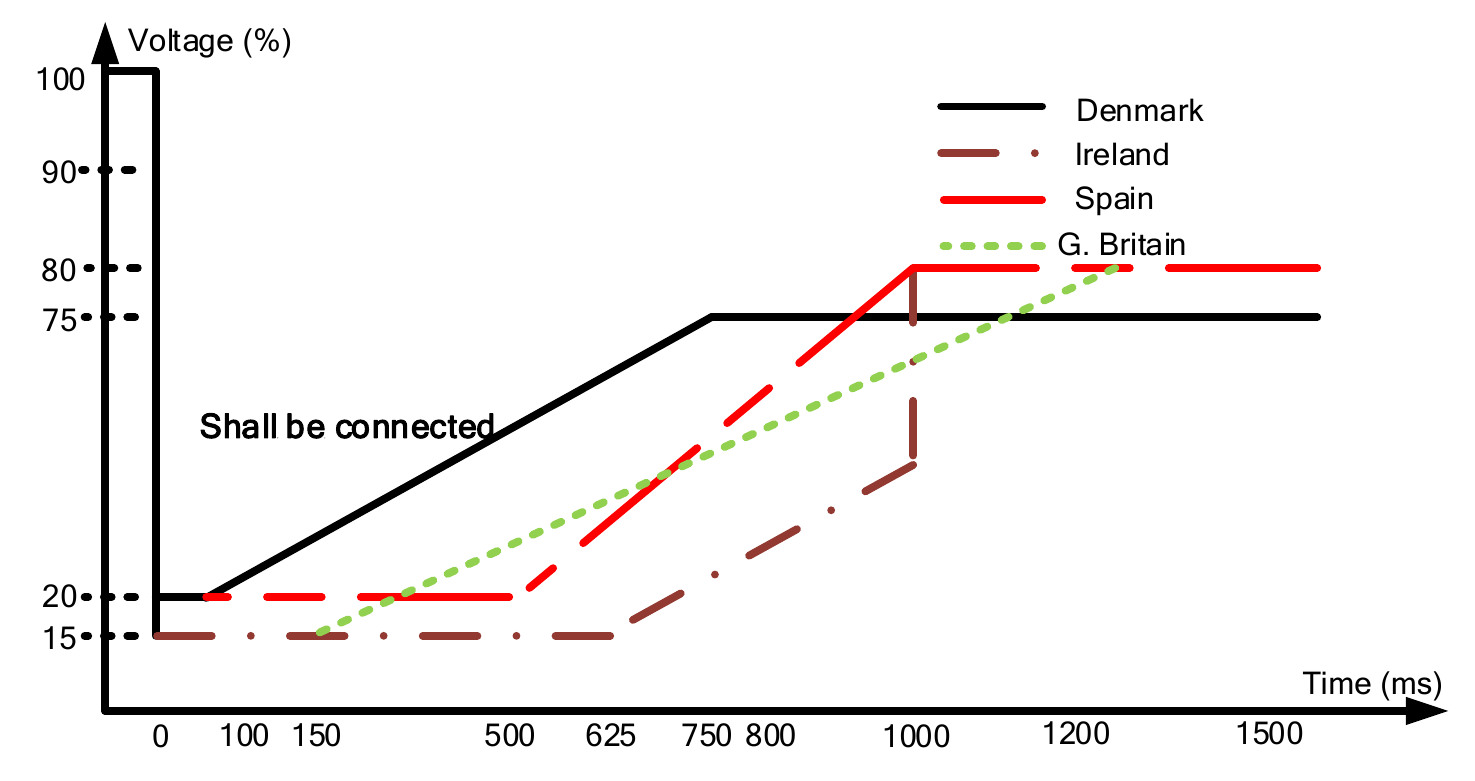}
\caption{Grid code for different countries}
\label{fig:grid_code}
\end{center}
\end{figure}

A quick disconnection of PV and wind plants, in case of disturbances, badly affects the stability of the system. Therefore, FRT capability requires that the PV/wind plants must remain connected to the grid during faults for a specific period. This requirement is mainly imposed by the modern grid code, which varies from country to country depending on different factors, \cite{ruano2007induction} as shown in Figure \ref{fig:grid_code}.

\begin{figure}[h]
	\begin{center}
		\includegraphics[scale=0.68]{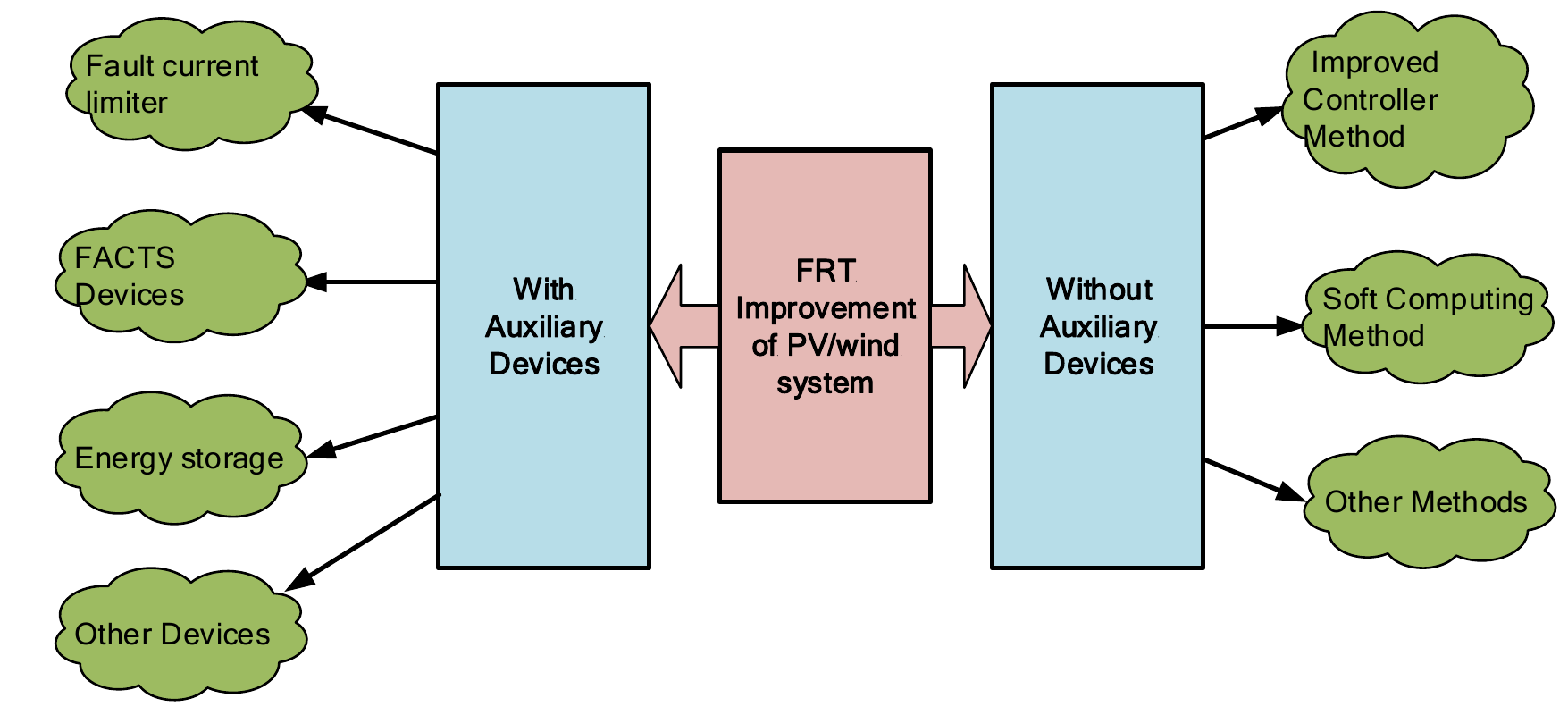}
		\caption{FRT improvement techniques of different renewable energy systems}
		\label{fig:frt_pv_wind}
	\end{center}
\end{figure}

Since the grid fault is responsible for low voltage of the system, it is important for the renewable sources to continue its operation to maintain continuity of the power flow and improve system reliability during contingencies. To achieve this goal, different improved control strategies are adopted and auxiliary devices are installed and controlled with the renewable energy sources \cite{alam2019fault,munkhchuluun2020long,el2013fault}. In the literature, FRT capability issues are discussed for three different system such as PV system, wind system, hybrid PV/wind system. Different techniques are presented for each of these systems as categorized in Figure \ref{fig:frt_pv_wind}. Mainly, FRT capability of PV/wind system is augmented without and with auxiliary devices. Improved control strategies and soft computing techniques are presented in the literature. In \cite{mohseni2011low}, a hybrid control technique is presented for improving both low voltage and high voltage FRT capability of DFIG wind system. A Sugeno fuzzy logic controller is presented in \cite{qais2020whale} which has less overshot and steady state error compared to the classical controller. In order to augment FRT of PV system, a model predictive strategy is presented in \cite{bighash2018improving} which has fast and robust control features. However, in  this method, additional controller is needed to switch between LVRT mode and normal mode which, in turn, increases cost of the controller. A detailed list of different control techniques without auxiliary devices is presented in Table \ref{tab:frt_without}. In the table, the advantages as well as different gaps in current study are well documented which can be a great source for the researchers for further study and improvement of FRT of renewable sources. 

\begin{figure}[H]
	\begin{center}
		\includegraphics[scale=0.68]{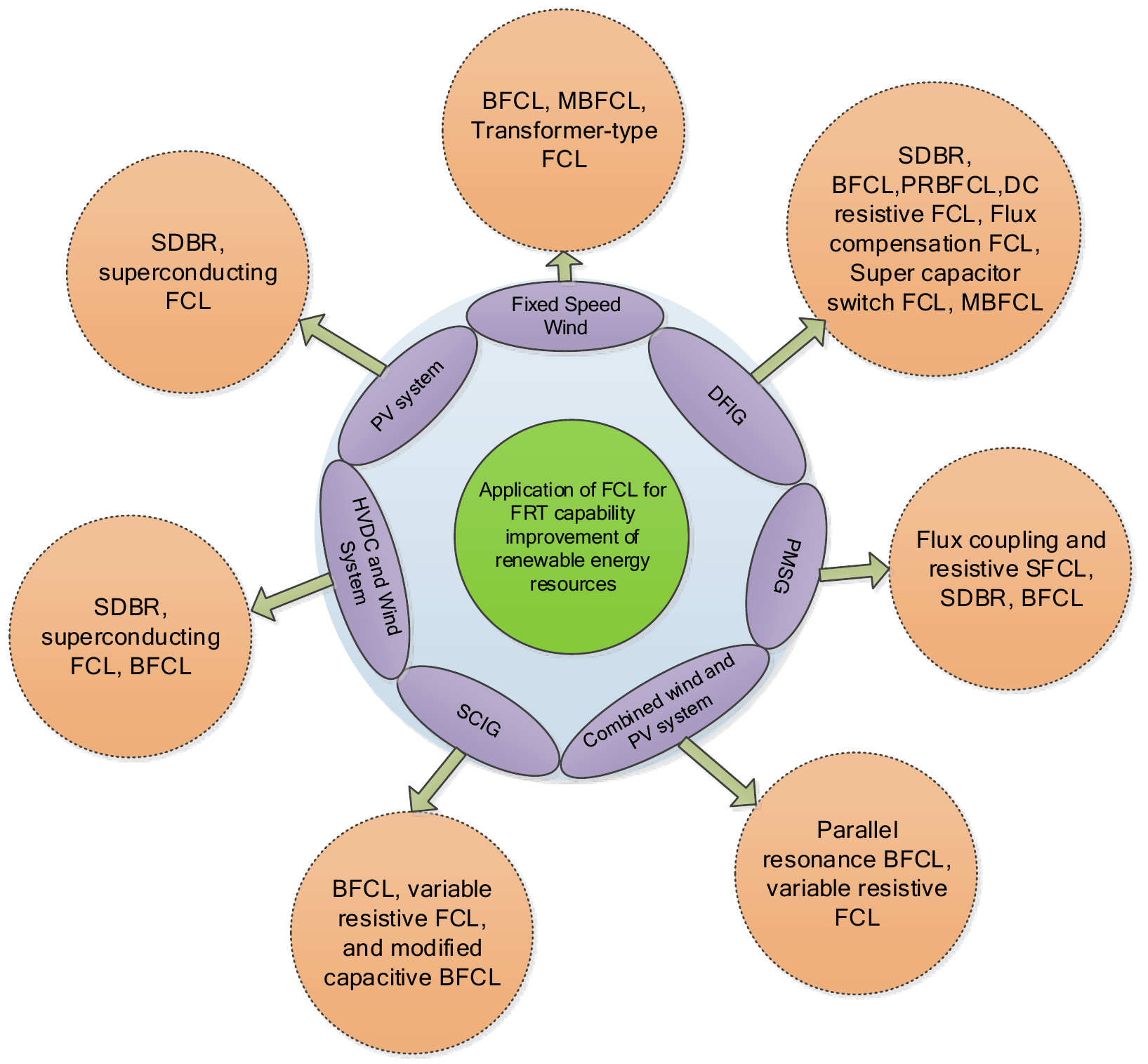}
		\caption{FRT improvement of renewable systems with FCLs}
		\label{fig:frt_fcl}
	\end{center}
\end{figure}

\begin{table}
	\centering
	\caption{FRT improvement techniques for different renewable energy systems}
	\label{tab:frt_without}
	\resizebox{\textwidth}{!}{%
		\begin{tabular}[H]{llll}
			\toprule
			Type [references] & Methods & Advantages & Disadvantages  \\[0.5ex] 
			\midrule    
			DFIG \cite{morshed2017new,benbouzid2014second,mohammadi2014efficient,amalorpavaraj2017improved,mohseni2011low}& Fuzzy based slide mode control & \tabitem Current and voltage within the limit & \tabitem Additional energy storage needed \\[0.2ex]
			& &\tabitem Continuation of DFIG operation &\\[0.2ex]
			& & \hspace{0.4cm} in non-ideal voltage case &\\[0.2ex]
			
			& Combined active and passive compensator method& \tabitem Reduced oscillation of DC link voltage and torque & \tabitem Increased cost and complexity \\[0.2ex]
			
			& Combined feed-forward and feed-back control& \tabitem Improved transient performance & \tabitem Dynamic voltage restorer (DVR) needed in grid side \\[0.2ex]
			
			& Hybrid control & \tabitem Improvement of both low and high voltage ride through & \tabitem PI parameters are not optimized \\[0.2ex]

			\midrule    
			PMSG \cite{chen2018reconfigurable,nasiri2016peak,jahanpour2019development,xing2017compositive,qais2020whale,yao2018coordinated} & Fault reconfigurable parallel control & \tabitem improved reliablity & \tabitem The system needs one additional controller \\[0.2ex] 
			&&\tabitem Multi-leg FRT capability&\tabitem Complex fault detection method \\
			& &\tabitem Only faulty leg is isolated &\\
			
			& Active power limitation controller&\tabitem Peak current within safe limit&\tabitem Not applicable for low inertia PMSG \\
			& &\tabitem DC link over voltage suppression & \tabitem Additional pitch angle controller needed\\
			& &\tabitem removal second-order active power fluctuation & \\
			
			& Combined vector and direct torque control&\tabitem Fast transient and smooth steady-state performance&\tabitem High cost due to use of two controllers \\
			& &\tabitem Reduction of rotor speed oscillation & \\
			
			& Composite control structure&\tabitem Reactive power support for grid voltage recovery&\tabitem Crowbar circuit is mandatory \\
			& &\tabitem Less stress on DC link capacitor &\tabitem Wind turbine must have 10\% over speed capability \\
			
			&  Sugeno fuzzy logic controller&\tabitem Quick response, less overshoot, negligible steady-state error&\tabitem High power loss in gearbox \\
			
			&  Coordinated controller&\tabitem Full use of each unit of a hybrid wind farm&\tabitem Complex and costly controller \\
			&  &\tabitem Improved stability and adaptability&\tabitem Real-time current and voltage measurements required \\
			&  &\tabitem No communication required among the wind farms&  \\
			
			\midrule  
			PV \cite{al2018low,bighash2018improving,huka2018comprehensive,mojallal2017enhancement,
				zheng2019design,wang2017fault,hasanien2015adaptive,wen2019new} & A novel LVRT control & \tabitem Protection of inverter during voltage dip & \tabitem A fast, automatic, and precise fault detection is mandatory for the controller \\[0.2ex] 
			&&\tabitem AC over-current and DC-link over voltage suppression &\tabitem A DC-chopper barker is needed to absorb energy \\
			
			&  Model predictive controller&\tabitem Fast and robust current control feature&\tabitem It needs additional controller for switching between normal mode and LVRT mode \\
			&  &\tabitem Low overshoot and fast tracking of reference signals&  \\
			
			&  Comprehensive LVRT strategy&\tabitem DC link over voltage reduction&\tabitem It can not track maximum power during fault  \\
			&  &\tabitem Converter over current reduction and higher reliability of PV system& \tabitem Complex control structure due to synthesis of positive and negative sequences  \\

			&  Nonlinear controller&\tabitem Improved recovery performance&\tabitem It needs two controller which increase cost  \\
			&  &\tabitem DC-link voltage remains within predefined limit during faults& \tabitem MPPT controller is switched off during fault  \\
			
			&  Model current predictive controller (MCPC)&\tabitem DC link harmonic reduction&\tabitem Complexity arises due to coordination between MCPC and non-MPPT controller  \\
			&  &\tabitem AC current is suppressed to preset value&   \\
			&  &\tabitem positive and negative sequence separation algorithm can all be removed&  \\
			
			& Robust control&\tabitem Improved DC bus voltage protection due to decoupling&\tabitem Estimation error of inductor current may cause unsatisfactory performance  \\
			&  &\tabitem Improved AC voltage profile&   \\
			
			& An adaptive control strategy&\tabitem PI controller has adaptive tuning feature  &\tabitem Initial assumption of PI parameter may deteriorate controller performance  \\
			&  &\tabitem DC link voltage fluctuation is reduced&   \\
			&  &\tabitem Power oscillation is well damped&   \\
			
			& Synchronous frame method&\tabitem Overcurrent protection for the converter  &\tabitem Unsatisfactory performance for deep voltage sag  \\
			&  &\tabitem No hard switching required between MPPT and non-MPPT controller& \tabitem Controller has negative impact on utility system   \\
			
			\midrule

			SCIG \cite{jiang2016combinational,naik2016improved,ahuja2016coordinated,jelani2014asymmetrical} & Combinational voltage booster technique & \tabitem Proper voltage is maintained during serious sag by controlling thyristor & \tabitem Complex control due to switching different operational modes \\[0.2ex] 
			&&\tabitem Excessive fault energy is dissipated in the braking resistor&  \\
			&&\tabitem Active power loss is minimized&  \\

			& Hybrid pitch angle controller &\tabitem Hybrid controller is superior over PI controller  &\tabitem High cost than PI due to extra controller \\
			&  &\tabitem Power and frequency oscillations are well damped&    \\
			
			& Coordinated FRT control &\tabitem Abnormal rise of DC link voltage is limited   &\tabitem  Double frequency oscillations in reactive power during unsymmetrical fault  \\
			&  &\tabitem Smoothing of injected reactive power during symmetrical fault&    \\
			
			& Distributed compensation controller &\tabitem Reduction of torque ripple   &\tabitem  For deeper voltage dip, higher current rating of constant power load is required  \\
			&  &\tabitem Improvement of wind farm reliability and stability& \tabitem Higher cost than centralized compensation controller    \\
			&  &\tabitem Independent control of positive and negative sequence voltages&    \\
					
			\midrule  
			Combined PV and Wind \cite{he2019coordinative,pagola2015low,morshed2019novel,noureldeen2017low} & Coordinative LVRT control & \tabitem Power imbalance reduction between faulted grid and renewable source & \tabitem Requirement of four controllers increases cost \\[0.2ex] 
			&&\tabitem Capable to handle transient voltage faults&\tabitem More stress on DC-link capacitor and rotating mass \\

			& Reactive power injection method &\tabitem PCC voltage profile is improved  &\tabitem PV array DC link voltage has overshoot \\
			&  &\tabitem The controller has feasibility to implement in hardware in loop&    \\

			& Auto-tuned fuzzy PI approach &\tabitem Less cost and simple design  &\tabitem Switching is needed for grid side converter controller \\
			&  &\tabitem Minimization rotor over current&  \\
			&  &\tabitem Voltage, power and torque fluctuation reduction&  \\

			& Modified controller &\tabitem Converter protection against over voltage  &\tabitem Switching among different controllers takes long time \\
			&  &\tabitem Reactive power support during faults&    \\
				
			\bottomrule
	\end{tabular}} \label{tab1}
\end{table}
 As shown in Figure \ref{fig:frt_pv_wind}, different auxiliary devices, such as fault current limier, energy storage, and FACTS devices, can be employed with RESs to augment FRT. The improvement of FRT for different PV and wind energy systems with the application of fault current limiters is shown in Figure \ref{fig:frt_fcl}.  Among the several auxiliary devices, fault current limiters (FCLs) are widely studied and implemented due to low cost, low loss in stand-by mode, high voltage withstanding capability. Mainly two types of FCLs are dominating for application in power system: superconducting and non-superconducting \cite{alam2018fault,alam2019model,liu2018impedance}.

 FRT enhancement of PV system, fixed speed wind, DFIG, PMSG, SCIG, wind integrated HVDC, and combined wind and PV systems are observed with different FCLs such as   bridge type fault current limiter (BFCL), series dynamic braking resistor (SDBR), modified BFCL, super conducting FCL, and variable resistive FCL \cite{huang2019cooperative,al2019fault}. 
As shown in Figure \ref{fig:frt_fcl}, FRT improvement of DFIG systems are achieved with almost all types of FCLs, whereas only few of them are studied and implemented in other renewable systems. This helps the novice researchers to find out the gaps in current study and fill-up those gaps with different cutting-edge technologies.  

An alternative solution to keep the renewable energy resources to be connected with the grid during disturbances is to use different FACTS devices such as static VAR compensator (SVC), thyristor controlled series capacitor (TCSC), and static synchronous compensator (STATCOM). In \cite{heydari2016novel}, a STATCOM based control strategy is proposed for the FRT improvement of fixed speed wind energy generation system. The hybrid PV/wind system FRT capability improvement is studied in \cite{ayvaz2016combined} with combined control of SVC and SDBR. In general, all these mentioned devices can help mitigate the fault problems of PV/wind system, however, integration of such devices increases both control complexity and cost. As an alternative to addition of the external devices, other methodologies are also presented in the literature to reduce cost and complexity. For instance, in \cite{benz2010low}, dynamic current limitation method is presented to augment FRT as well as save inverter of a small-scale solar system. It is worth mentioning that modification or improvement of such method is needed for large-scale PV system or hybrid PV/wind systems. This gap, FRT improvement of large-scale hybrid wind/PV system, could be filled-up by cutting-edge technologies.

Another approach of improving FRT of RESs is to use energy storage systems (ESSs) such as battery, supercapacitor, and fly wheel energy storage. The main function of the energy storage is to absorb energy from the system during disturbances so that the negative impact of faults is minimized.  In \cite{saadat2014statistical}, capacitor energy storage system is investigated for FRT improvement of distributed renewable generator. Since the supercapacitor has high power density, it is proposed as a potential solution to reduce short term power fluctuation in PV system during normal condition \cite{worku2015grid}. Also, during the grid side faults, engery generated by PV is stored in the supercapacitor to augment FRT capability of PV system. In general, energy storage is a costly solution to the FRT problems of renewable energy system. Further investigation is needed to minimize the cost by optimal sizing of ESSs as well as combined FCL and ESSs could be a better solution for FRT of PV/wind system.

\section{ Power quality issues} \label{secharmonics}

\begin{figure}[h]
	\begin{center}
		\includegraphics[scale=0.78]{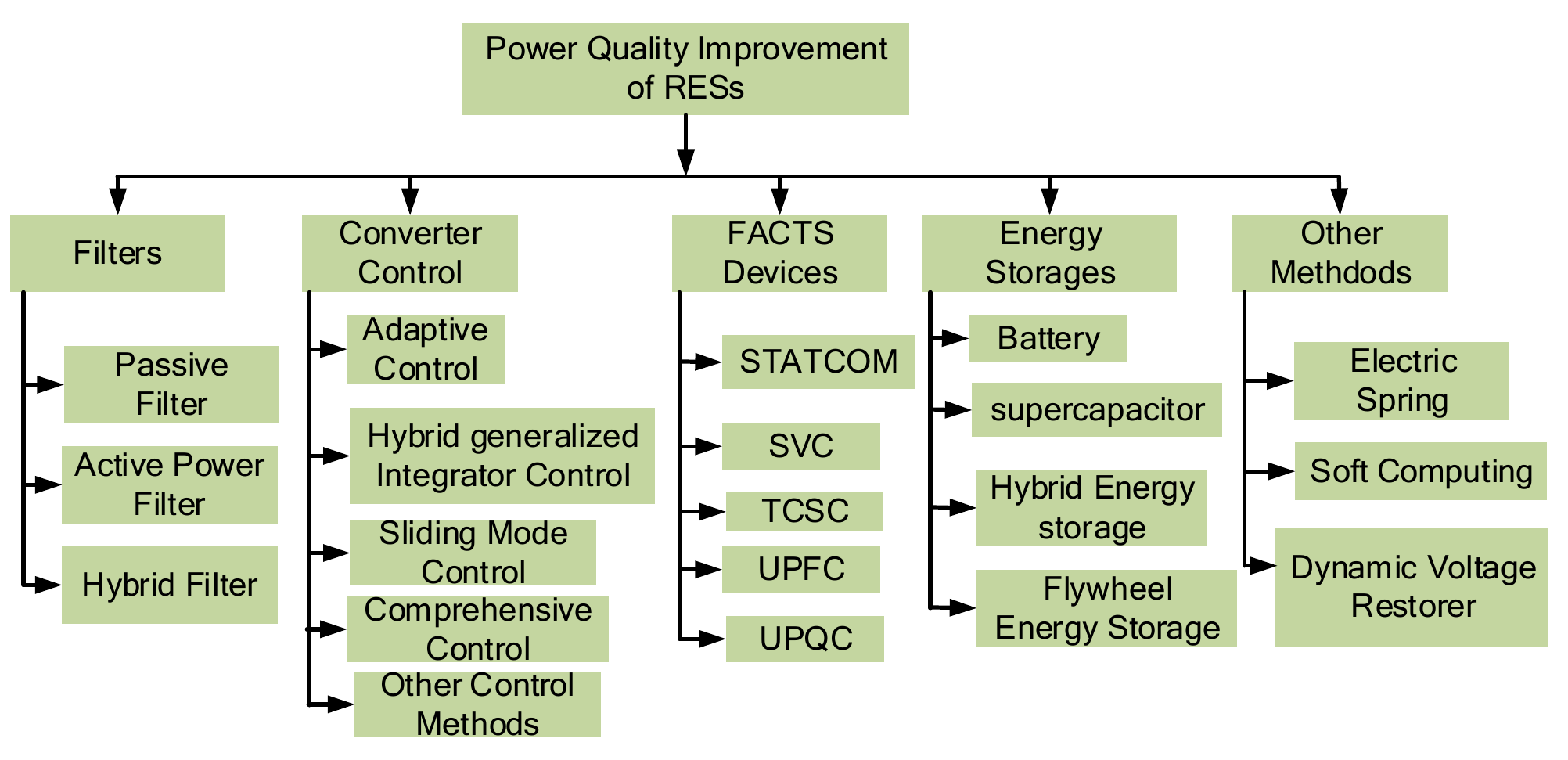}
		\caption{ Power quality improvement techniques for RESs}
		\label{fig:power_quality}
	\end{center}
\end{figure}

The heart of  renewable energy system is the power electronic (PE) converters. These devices are responsible for the harmonic injection in the system. Furthermore, operation of these converters are highly dependent on the quality of the voltage signal. In order to improve the power quality of the RESs different measures are taken such as implementation of improved control strategies and use of different auxiliary devices \cite{liang2018harmonics}.  The different approaches and cutting-edge technologies used for power quality improvement are visualized in Figure \ref{fig:power_quality}.

As shown in Figure \ref{fig:power_quality}, filters, flexible AC transmission systems (FACTS) devices, energy storages, and converter control can be sub-categorized into different ones in order to improve power quality. Some of these methods are also documented in the previous sections for reducing power fluctuation and frequency deviation. Power quality issues of PV/wind systems can be handled with more advanced filtering technologies such as active and passive filters \cite{djeghloud2011sub,ishaque2013review,elkholy2018harmonic}.  However, the cost, size and weight of passive filters (PFs) increase with the increase of power rating of the converters. Thus, PFs are not a better solution for cutting-edge technology \cite{tali2014passive,tareen2017active}. On the other hand, one of the most attractive solutions is to employ the shunt active power filter (SAPF) in order to improve power quality, reactive power and current harmonics compensation \cite{ravinder2019investigations,kolar2011review}. With the increase of the penetration of renewable energy, the active power filter size increases. In order to resolve this issue and minimize the filter size and cost, hybrid APF is proposed \cite{yamada2010starting,litran2012analysis}. In hybrid filter, lower order harmonics are eliminated by SAPF whereas the PF removes the higher order harmonics \cite{hui2018efficient,qian2008improved}. The harmonics at switching frequency and the multiple of switching frequency are problematic for the converters of renewable energy system. In order to mitigate these harmonics with less voltage drop and small component size, another higher order active filter is presented in \cite{anzalchi2017design}.  

The different advanced control methodologies of converter of renewable energy system are presented in order for harmonic mitigation \cite{chishti2019weak,yu2018dse,chishti2019lmmn,bubshait2017power}. An improved high frequency harmonics rejection technique is presented in \cite{chishti2019weak} with hybrid generalized integrator controller. Although the controller has a trade-off between accuracy and speed of convergence, it eliminates both interharmonics, subharmonics and disturbances. In \cite{yu2018dse}, a dynamic state estimation based slide mode control is proposed for grid connected DFIG wind farm, which is capable to alleviate unnecessary switching of converters as well as improve power quality. Some control techniques are proposed for the hybrid PV/wind systems in \cite{chishti2019lmmn,seghir2018new}, for the improvement of power quality, without any filter or auxiliary devices. As reported in \cite{chishti2019lmmn}, harmonics compensation and fundamental load
component extraction are achieved with a new least mean mixed norm (LMMN) control strategy.

The flexible AC transmission systems (FACTS) devices play an important role to improve different aspects of power quality, like harmonics, power factor, oscillations in electrical quantities, voltage dip, in highly renewable penetrated systems \cite{elmetwaly2020adaptive, gandoman2018review,abdelsalam2012novel}. Various FACTS devices, such thyristor controlled series capacitor (TCSC), static var compensator (SVC), and static synchronous compensator (STATCOM), are presented in the literature to handle harmonics issues of renewable energy system. A detail documentations on harmonic mitigation with different FACTS devices are visualized in table \ref{tab:facts}. Several gaps on different approaches are clearly mentioned which can be a good source for future research.

\begin{table}[htbp]
	\centering
	\caption{FACTS devices for power quality improvement}
	\resizebox{\textwidth}{!}{%
	\begin{tabular}{lll}
		\hline
		\multicolumn{1}{l}{\textbf{FACTS Devices}} &
		\multicolumn{1}{l}{\textbf{Contirbutions}} &
		\multicolumn{1}{l}{\textbf{Comments}}
		\bigstrut\\
		\hline
			
			D-FACTS \cite{elmetwaly2020adaptive,jyotishi2013mitigate} & \tabitem These methods improve
		dynamic voltage stabilization & \tabitem Further improvement is possible by systematic D-FACTS design
		\bigstrut\\ 
	& \tabitem Harmonics mitigation is achieved and power factor is improved &
		
		\bigstrut\\
		\hline
		
		SSFC \cite{abdelsalam2012novel} & \tabitem Harmonics mitigation and voltage stabilization & 
				\bigstrut\\
		
		 & \tabitem Losses reduction and power factor improvement & 
		\bigstrut\\
		\hline
		
		STATCOM  \cite{sunil2012power,chavan2015using,jamil2019power} & \tabitem Point of common coupling harmonic mitigation & \tabitem Voltage controller parameters can be tuned with any optimization technique
		\bigstrut\\ 
		& \tabitem Voltage deviaton reduced by 73.4\% in \cite{jamil2019power} & \tabitem The strategy presented in \cite{jamil2019power} only considers linear load
		\bigstrut\\
		\hline
		TCSC\cite{thampatty2019design,kuang2018voltage}& \tabitem Machine electrical torque deviation is reduced &  \tabitem In some cases, damping of oscillation is slower than traditional controller \cite{thampatty2019design}
		\bigstrut\\
		& & \tabitem Optimal placement of TCSC is not considered \cite{kuang2018voltage}
		\bigstrut\\
		\hline
		SVC\cite{savic2014optimal,tayyebifar2014performance}& \tabitem Wind and PV systems voltage profile is improved   &  \tabitem Harmonics minimization is not considered in optimization \cite{savic2014optimal} 
		\bigstrut\\
		& \tabitem Large-scale DFIG harmonic reduction is observed in \cite{tayyebifar2014performance}&
		\bigstrut\\
		\hline
		UPFC \cite{panah2016reactive,bhargava2019power,kalair2017review}&\tabitem Voltage profile and harmonics are improved &\tabitem UPFC can mitiage more harmonics with phase shifting capability \cite{kalair2017review}
		\bigstrut\\
			\hline
			UPQC \cite{hossain2018analysis,bhavani2015fuzzy,chaudhary2016enhancement,rashad2018performance}&\tabitem Voltage sag and current harmonics are mitigated & \tabitem Mitigation of higher order harmonics requires further research
			\bigstrut\\
			& \tabitem Interharmonics, noise and DC offset are improved&
			\bigstrut\\
			\hline
	\end{tabular}}%
	\label{tab:facts}%
\end{table}%

Different energy storage devices, such as battery, supercapacitor, are flywheel energy storage, are employed to improve power quality of renewable energy system, especially for the purpose of power smoothing \cite{liu2011cascaded,behravesh2019control,djamel2013power,seo2010power,worku2019fault}. A sophisticated power allocation method between PV and battery storage is developed in \cite{liu2011cascaded} to mitigate over voltage at PCC and support wide range of reactive power.  Improvement of power quality for a PV system is observed in \cite{djamel2013power} with battery storage which is connected with the DC link of the PV energy conversion system. However, the battery storage is not suitable for frequent charging and discharging applications due to its low power density and small life cycle. In order to resolve this issue, a hybrid energy storage, battery and supercapacitor, is proposed in \cite{ding2017novel} to smooth the power fluctuation in a wind energy integrated system. Both high energy density of battery and high power density of supercapacitor are harnessed in this new approach. This work mainly employs self-adaptive wavelet packet decomposition and two level power reference signal distribution technique to reduce grid power fluctuation due to wind speed variations. The parameters of battery and supercapacitor are given by experience in this study; however, the economic optimization can be employed for further power quality improvement. Power quality improvement of PV and DFIG systems is presented  \cite{seo2011power,mukherjee2020effective} with superconducting magnetic energy storage (SMES). The high temperature superconducting (HTS) coil is charged and discharged based on PV array output and utility power quality \cite{seo2011power}. Although the power quality of DFIG wind system is improved in \cite{mukherjee2020effective}, extreme high current flows through the SMES coil which may be problematic for practical implementation of such device. To resolve this issue, improved control strategy can be proposed or several SMES can be connected in parallel as a future research. Moreover, the power quality improvement of renewable energy systems (RESs) is also achieved with some other methodologies like electric spring, dynamic voltage restorer (DVR), soft computing based methods\cite{soni2017electric,benali2018power,latif2020state,amjad2014review}. A hybrid PV/DFIG system harmonics mitigation is presented in \cite{benali2018power} with a fuzzy logic controlled DVR. However, the proposed technique does not take the voltage deviation at the PCC and harmonic contents of voltage signals as input to fuzzy controller. Further harmonics improvement may be possible with the consideration of these facts.       

\section{Uncertainty Issues}

\begin{figure}[h]
	\begin{center}
		\includegraphics[scale=0.78]{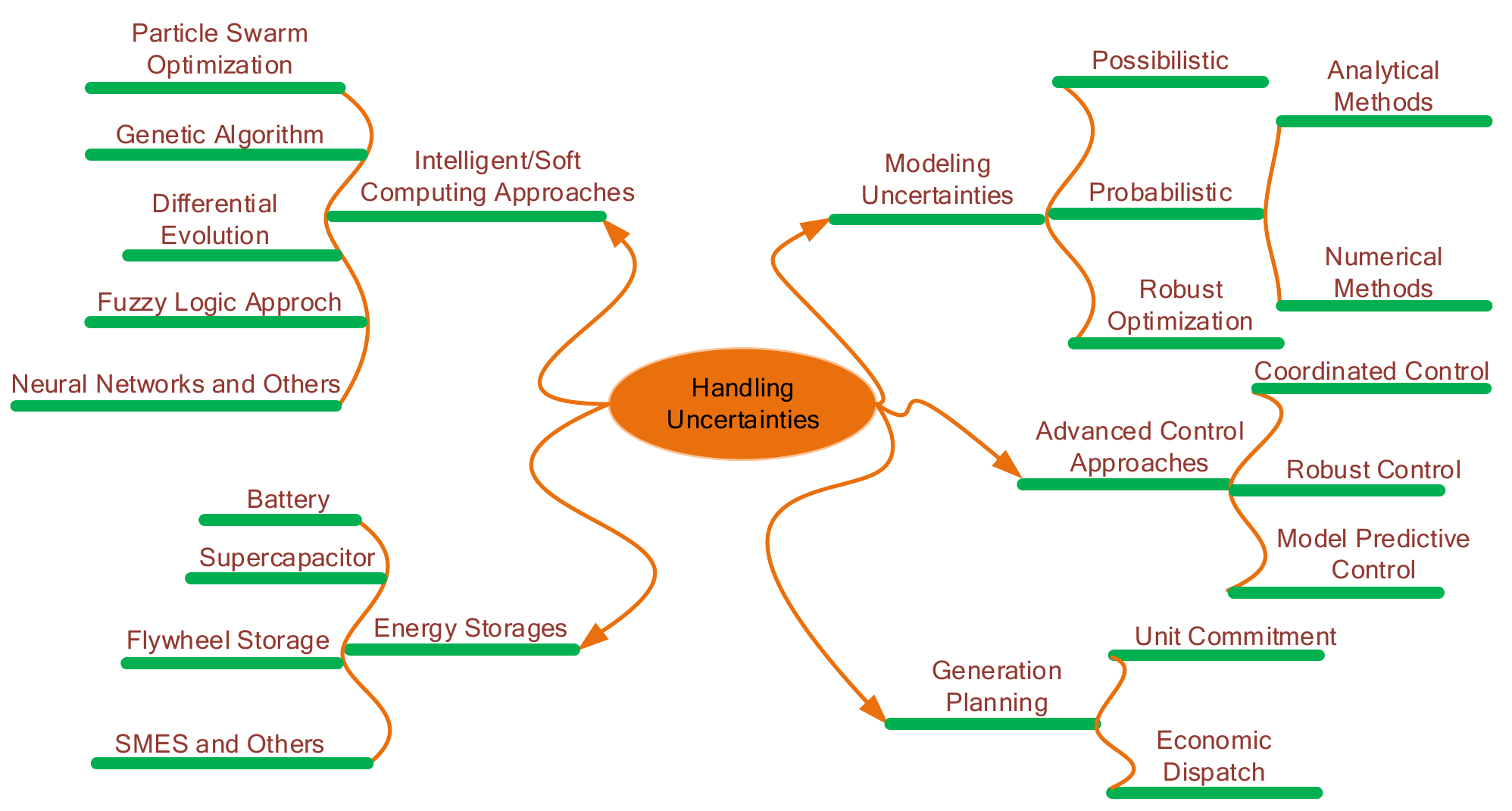}
		\caption{ Pictorial elaboration of uncertainty mitigation methods }
		\label{fig:uncer}
	\end{center}
\end{figure}

In power systems, uncertainty means inaccurate parameters which can not be predicted with 100\% certainty and which affects the smooth operation. 
Nowadays, renewable energy sources are considered as main source of uncertainty in power system due to their intermittent nature \cite{wang2017efficient}.
With the high-level integration of intermittent renewable sources to the grid, the main question remain; how do the system operators manage the uncertainty from these sources? However, vast majority of the optimization techniques, soft computing and advanced control algorithms, energy storage devices are employed to mitigate the uncertainty issues. Numerous methods are implemented to fully or partially mitigate the uncertainties in RESs integration. Of them, the key approaches are shown in Figure \ref{fig:uncer} and summarized below.    

\begin{itemize}
	
\item \textbf{Modeling Uncertainties:} 

 The probabilistic pattern of wind and solar power generation systems are caused by different operational challenges, which stem from uncertainty in weather, wind speed, and solar irradiation \cite{baghaee2017application}. The uncertainty in loads, correlated wind, solar distributed energy resources, and plug-in hybrid electric vehicles are modeled in \cite{baghaee2017fuzzy} with possibilitic method. The uncertainty modeling of wind energy conversion adopts dynamic model which is further partitioned into stochastic and deterministic components\cite{zakaria2020uncertainty}. A dynamic empirical wind turbine power curve (WTPC) model is developed based on Langevin model and maximum principle method \cite{gottschall2008improve}. Another probabilistic WTPC model, based on normal distribution, varying mean and constant standard deviation, is proposed in \cite{jin2010uncertainty} to minimize uncertainty. Hitherto, most of the mentioned methods simulate WTPC uncertainty based on known distribution and statistical parameters, which may not be consistent with real situation. Furthermore, the evaluation of probabilistic model is more complicated than deterministic one, thus it requires a new evaluation criteria. To address these issues, a probabilistic WTPC model is proposed \cite{yan2019uncertainty} with new model inputs, such as pitch angle and wind direction, and evaluation criteria,  to quantify the uncertainties of energy conversion. The uncertainty modeling of a distribution network, comprising solar and wind generation systems, is presented \cite{shang2019equivalent} in which deterministic and uncertain components are calculated based on fitted power characteristic and probability distribution, respectively. The uncertainty management to assist high-level RESs integration process is also achieved with robust optimization techniques as presented in the literature \cite{verastegui2019adaptive,tian2019coordinated,yi2015robust}. There is extreme need for new long-term planning models of power system to incorporate uncertainty resulted from large scale renewable energy integration. In order to address this issue, in \cite{verastegui2019adaptive}, a new generation and transmission expansion planning model is proposed based on robust optimization. The key feature of the model is that daily uncertainty is represented by the concept of uncertainty sets of representative days, which measures loads and renewable generations over such days.

\item \textbf{Generation Planning:} Another approach to alleviate the risk of uncertainty is to implement improved and intelligent generation planning/scheduling with the help of unit commitment (UC) and economic dispatch (ED). A comparative study of renewable uncertainty integration into stochastic security-constrained unit commitment is performed in \cite{quan2016integration}. Both stochastic and deterministic models are studied where the stochastic model demonstrates more robust performance. Due to the fact that the gas-fired generating units are capable of ramping up quickly, which can mitigate uncertainty of renewable energy, many countries are installing more gas-fired units \cite{devlin2016importance}. Higher the uncertainty in renewable energy, higher the dependency of the power system on traditional gas-fired generation, which transfers the uncertainties from renewable source to gas flow network.  Thus, it necessitates an integrated unit commitment problem for electricity network and gas network. To come up with this issue, a new integrated unit commitment problem is formulated for electric and gas networks and is solved with mixed integer non-linear optimization technique in \cite{fallahi2020integrated}. Most of the studies mainly focus on one or two aspects of uncertainties; however, inclusion of all possible uncertainties is required for better system operation. Some  unit commitment methodologies are presented in     \cite{melamed2018multi,quan2015computational,li2019network}, dealing with the mitigation process of both forecast and outages uncertainties of solar and wind power generation. Another new multiobjective unit commitment optimization technique, utilizing information-gap decision theory (IGDT), is proposed in \cite{ahmadi2019information} considering both wind power and load demand uncertainties. The impact of variability of wind and solar power can also be minimized with economic dispatch (ED), short-term optimal output power scheduling from a number of generation units to meet the load demand subject to network and other constraints. A model, through Weibull distribution evaluated on a linearized  power curve of the wind farm, is developed in \cite{hetzer2008economic} to solve ED problem considering wind power generation uncertainty. A similar model is proposed \cite{liu2010economic} in which wind generation is considered as constraint in optimization problem and is analyzed with Lagrange multiplier approach. A combined UC and ED solution method, which is described by the probability distribution function of thermal generators' output power, excess electricity, energy not served, and spinning reserve, is reported in \cite{lujano2016new} to reduce uncertainty impacts of high-level RESs integration. In this method, combination of priority list and ED solution is used to solve UC problems. Another unified UC and ED is solved \cite{bakirtzis2018storage} for RESs system having energy storage devices for short-term operation scheduling. This work may be further extended to include more electricity generation mix, which includes more gas-fired units and renewable sources. 

\item \textbf{Advanced Control Approaches:} Several control approaches  are also presented in the literature to handle the uncertainties stem from the variation of wind speed and solar irradiation \cite{wang2013frequency,ochoa2018frequency}. The presented control approach in \cite{ochoa2018frequency} combines the virtual inertia concept and pitch angle control to dynamically shift the maximum power tracking curve of DFIG wind generation system. The output power smoothing  for wind generation systems, DFIG and PMSG, are presented in \cite{lyu2018coordinated,ouyang2019multi,jiang2012two} with coordinated control approaches. The voltage control, rotor speed control, and pitch angle control are coordinated in hierarchical manner to reduce the impact of uncertainties  due to wind speed variation \cite{lyu2018coordinated}. Most of the control approaches do not deal with multi disruption and unknown parameters. However, the model predictive control (MPC) approach has the ability to consider both of them. In order to take these advantages, MPC  based control approaches are presented in \cite{ouyang2020multi,liu2018model} for uncertainty mitigation of wind and solar energy systems.   A robust nonlinear controller is presented in  \cite{mahmud2014robust} for grid connected solar PV system. The partial feedback linearization approach is adopted and the robustness is guaranteed with wide range of uncertainties in solar system.

\item \textbf{Intelligent/Soft Computing Approaches:} The reliability of network performance is degraded due to large scale stochastic renewable energy integration to the grid. Soft computing approaches are extensively employed in power system for reduction  uncertainty impact resulted from stochastic renewable energy sources. In \cite{mozafar2017simultaneous}, genetic algorithm-particle swarm optimization (GA-PSO) based hybrid technique is reported for wind and solar system. The load uncertainty and random change of demand are considered in \cite{gholami2019modified} in optimization problem which is solved by modified PSO. Due to the efficacy to solve the optimization problem with many operational constraints, differential evolution (DE) is reported in \cite{bu2019species,awad2019efficient} for the renewable system in which generation uncertainties are on the top of load variations. In order to minimize the renewable energy prediction, artificial neural network (ANN) is trained with uncertain parameters like solar irradiance and wind speed \cite{masoumi2020application}. Afterwords, eco-static objective function, reliability criteria, and battery management strategy are modeled and optimized with weighted improved PSO algorithm for further minimization of impact of uncertainty.

\item \textbf{Energy Storages:} 
 In RESs integration, uncertainty resulted from  unavailability of exact information causes several unexpected system behaviors. As mentioned before, several optimization/soft computing approaches enable the development of RES models that are resilient to uncertainties; however, implicit considerations are required to build such models in case of intense presence of uncertainties. Still, the RES model needs to be designed systematically considering its high-level inherently fluctuating phenomena. To address this issue, several storage devices are the best candidates to be integrated within the system model\cite{hakimi2016smart,giannakoudis2010optimum}. A systematic design approach to include highest level of flexibility in RES model is proposed considering hourly demand response, energy storage devices, and fast ramping unit \cite{nikoobakht2018assessing}. To harness the energy-shifting and fast ramping capabilities of batteries, stochastic unit commitment and energy scheduling with economic dispatch are presented \cite{li2015flexible}. Further investigation on comprehensive set of real time operation strategies of batteries could be new research direction and or more detailed life-time degradation impact of batteries on uncertainty of RESs can be investigated. The source and load uncertainties management scheme is developed in \cite{akram2019hybrid} considering a hybrid energy storage system, comprising battery and supercapacitor. The performance of this hybrid scheme is also evaluated with other possible hybrid schemes, like superconducting magnetic energy storage and batteries (SMES-batteries) and flywheel-batteries. As per present discussions, energy storage system (ESS) plays an important role in mitigating and managing uncertainties; however, determination of size of ESS related  to uncertainty mitigation process is imperative.  A theoretical ESS sizing method is proposed in \cite{oh2020theoretical} considering stochastic nature of uncertainties and analyzed by mean absolute error (MAE). Different approaches, probabilistic \cite{wu2014statistical}, optimization \cite{yang2014joint}, frequency domain \cite{oh2018energy}, in evaluating the size of ESS are also adopted for mitigation of uncertainties in RESs integration.

\end{itemize}

\section{Current Challenges and Future Recommendations}

Nowadays, several economic and technical benefits are gained from high-level integration of converter based renewable energy sources, such as low cost energy, less carbon emission, less operational and maintenance cost. However, many  technical issues, very low inertia causing frequency instability, high fault current resulting from the short circuit, intensive uncertainties due to varying nature of wind speed and irradiance, degraded power quality, are raised due to high-level renewable integration. It is challenging for the researchers, system planners and operators to maintain flexible, reliable, and stable operations of such systems. However, several cutting-edge technologies, techniques are used and being continuously developed to deal with new challenges resulted from renewable energy integration. For instance, virtual inertia control, fault current limiters, advanced filters, energy storage devices, optimization techniques are adopted to handle those issues. Nevertheless, as still there are possibilities to develop better strategies to deal with the several challenges of high-level renewable energy integration, the future researches are likely to be conducted in the areas summarized bellow.   
 
\begin{itemize}
\item Advanced control methodologies (such as predictive, adaptive, intelligence, robust, optimal,  hierarchical control) can be redesigned/improved/implemented considering high power rating of converters and improved power sharing among the converters in order to facilitate high-level RES integration. Nevertheless, efficient power sharing of RES is often overlooked with the ideal voltage source assumption to the input of the converters.

\item Even with the advanced control strategies, RES system is still vulnerable to the faults. Some auxiliary devices, such as recently developed non-superconducting fault current limiters, can be analyzed, and applied to RES system, and their feasibility studies can be conducted.       
\item Application of advanced control strategy highly depends of proper modeling of the system. Thus, improved and or novel model can be developed considering stochastic nature of renewable sources. Furthermore, importance should be given to reduce the complexity of such models in order for easy practical implementation.	

\item Low inertia is the serious concern for high-level RES integration. Although some methodologies, such as virtual inertia controllers, and droop controller, are present in the literature to ease the problem, still there are opportunities to contribute in this direction by designing improved inertia controller. Specially, the amount of virtual inertia needed for stable operation of high RES system can be optimized with advanced algorithms, and then this amount could be supported with improved virtual controller.  

\item There are few research studies that deal with the inertia/frequency support of RES system without auxiliary system. The DC side of converters of RES consists  capacitors and AC side consists inductors. These two elements have build-in energy storing capability and   mimic the  behavior of rotating mass. Thus, voltage and current control of these elements can be hot research topic for short-term frequency support from them, similar to the frequency support from rotating mass inertia of a synchronous generator.     

\item Energy storage devices are important to support frequency, and voltage profile. However, few researches consider the impact of lifetime of energy storages when they are applied for frequency and voltage support, power quality improvement, and uncertainty management.

\end{itemize}

\section{Conclusion}  \label{SecCon}

Although high-level RES integration to the grid has advantages, it still raises several serious concerns like low inertia, degraded power quality, and high-level uncertainties. The aim of this article is to offer a comprehensive and in-depth review on challenges and solutions of high-level RES integration  to the grid. Several challenges, such as low inertia, high fault current, low power quality, high uncertainty, are clearly pointed out and discussed. Besides, potential solution to each challenge is well documented  with recent research articles. The detailed analysis of several technical problems of high-level RES integration are well documented with necessary graphical representations.  Several gaps in current study are clearly pointed out as challenges which can be filled up with cutting-edge technologies and novel researches. It is expected that this article can be an excellent reference for researchers of all level, beginner to high level, to realize major challenges and opportunities in RES integration to the gird. Finally, a complete list of future works are highlighted for further improvement in control and operation of renewable energy systems.

\bibliography{review_ref}

\end{document}